\documentclass[
 reprint,
 superscriptaddress,
 amsmath,amssymb,
 aps, prd,
]{revtex4-2}[prd]

\usepackage{graphicx}
\usepackage{wrapfig}
\usepackage{xcolor}
\usepackage[caption=false]{subfig}
\usepackage{dcolumn}
\usepackage{bm}
\usepackage{hyperref}

\begin{document}

\preprint{TODO}

\title{Physics potential of the IceCube Upgrade for
atmospheric neutrino oscillations}

\affiliation{III. Physikalisches Institut, RWTH Aachen University, D-52056 Aachen, Germany}
\affiliation{Department of Physics, University of Adelaide, Adelaide, 5005, Australia}
\affiliation{Dept. of Physics and Astronomy, University of Alaska Anchorage, 3211 Providence Dr., Anchorage, AK 99508, USA}
\affiliation{School of Physics and Center for Relativistic Astrophysics, Georgia Institute of Technology, Atlanta, GA 30332, USA}
\affiliation{Dept. of Physics, Southern University, Baton Rouge, LA 70813, USA}
\affiliation{Dept. of Physics, University of California, Berkeley, CA 94720, USA}
\affiliation{Lawrence Berkeley National Laboratory, Berkeley, CA 94720, USA}
\affiliation{Institut f{\"u}r Physik, Humboldt-Universit{\"a}t zu Berlin, D-12489 Berlin, Germany}
\affiliation{Fakult{\"a}t f{\"u}r Physik {\&} Astronomie, Ruhr-Universit{\"a}t Bochum, D-44780 Bochum, Germany}
\affiliation{Universit{\'e} Libre de Bruxelles, Science Faculty CP230, B-1050 Brussels, Belgium}
\affiliation{Vrije Universiteit Brussel (VUB), Dienst ELEM, B-1050 Brussels, Belgium}
\affiliation{Dept. of Physics, Simon Fraser University, Burnaby, BC V5A 1S6, Canada}
\affiliation{Department of Physics and Laboratory for Particle Physics and Cosmology, Harvard University, Cambridge, MA 02138, USA}
\affiliation{Dept. of Physics, Massachusetts Institute of Technology, Cambridge, MA 02139, USA}
\affiliation{Dept. of Physics and The International Center for Hadron Astrophysics, Chiba University, Chiba 263-8522, Japan}
\affiliation{Department of Physics, Loyola University Chicago, Chicago, IL 60660, USA}
\affiliation{Dept. of Physics and Astronomy, University of Canterbury, Private Bag 4800, Christchurch, New Zealand}
\affiliation{Dept. of Physics, University of Maryland, College Park, MD 20742, USA}
\affiliation{Dept. of Astronomy, Ohio State University, Columbus, OH 43210, USA}
\affiliation{Dept. of Physics and Center for Cosmology and Astro-Particle Physics, Ohio State University, Columbus, OH 43210, USA}
\affiliation{Niels Bohr Institute, University of Copenhagen, DK-2100 Copenhagen, Denmark}
\affiliation{Dept. of Physics, TU Dortmund University, D-44221 Dortmund, Germany}
\affiliation{Dept. of Physics and Astronomy, Michigan State University, East Lansing, MI 48824, USA}
\affiliation{Dept. of Physics, University of Alberta, Edmonton, Alberta, T6G 2E1, Canada}
\affiliation{Erlangen Centre for Astroparticle Physics, Friedrich-Alexander-Universit{\"a}t Erlangen-N{\"u}rnberg, D-91058 Erlangen, Germany}
\affiliation{Physik-department, Technische Universit{\"a}t M{\"u}nchen, D-85748 Garching, Germany}
\affiliation{D{\'e}partement de physique nucl{\'e}aire et corpusculaire, Universit{\'e} de Gen{\`e}ve, CH-1211 Gen{\`e}ve, Switzerland}
\affiliation{Dept. of Physics and Astronomy, University of Gent, B-9000 Gent, Belgium}
\affiliation{Dept. of Physics and Astronomy, University of California, Irvine, CA 92697, USA}
\affiliation{Karlsruhe Institute of Technology, Institute for Astroparticle Physics, D-76021 Karlsruhe, Germany}
\affiliation{Karlsruhe Institute of Technology, Institute of Experimental Particle Physics, D-76021 Karlsruhe, Germany}
\affiliation{Dept. of Physics, Engineering Physics, and Astronomy, Queen's University, Kingston, ON K7L 3N6, Canada}
\affiliation{Department of Physics {\&} Astronomy, University of Nevada, Las Vegas, NV 89154, USA}
\affiliation{Nevada Center for Astrophysics, University of Nevada, Las Vegas, NV 89154, USA}
\affiliation{Dept. of Physics and Astronomy, University of Kansas, Lawrence, KS 66045, USA}
\affiliation{Dept. of Physics, King's College London, London WC2R 2LS, UK}
\affiliation{Centre for Cosmology, Particle Physics and Phenomenology - CP3, Universit{\'e} catholique de Louvain, Louvain-la-Neuve, Belgium}
\affiliation{Department of Physics, Mercer University, Macon, GA 31207-0001, USA}
\affiliation{Dept. of Astronomy, University of Wisconsin{\textemdash}Madison, Madison, WI 53706, USA}
\affiliation{Dept. of Physics and Wisconsin IceCube Particle Astrophysics Center, University of Wisconsin{\textemdash}Madison, Madison, WI 53706, USA}
\affiliation{Institute of Physics, University of Mainz, Staudinger Weg 7, D-55099 Mainz, Germany}
\affiliation{Department of Physics, Marquette University, Milwaukee, WI 53201, USA}
\affiliation{Institut f{\"u}r Kernphysik, Universit{\"a}t M{\"u}nster, D-48149 M{\"u}nster, Germany}
\affiliation{Bartol Research Institute and Dept. of Physics and Astronomy, University of Delaware, Newark, DE 19716, USA}
\affiliation{Dept. of Physics, Yale University, New Haven, CT 06520, USA}
\affiliation{Columbia Astrophysics and Nevis Laboratories, Columbia University, New York, NY 10027, USA}
\affiliation{Dept. of Physics, University of Oxford, Parks Road, Oxford OX1 3PU, United Kingdom}
\affiliation{Dipartimento di Fisica e Astronomia Galileo Galilei, Universit{\`a} Degli Studi di Padova, I-35122 Padova PD, Italy}
\affiliation{Dept. of Physics, Drexel University, 3141 Chestnut Street, Philadelphia, PA 19104, USA}
\affiliation{Physics Department, South Dakota School of Mines and Technology, Rapid City, SD 57701, USA}
\affiliation{Dept. of Physics, University of Wisconsin, River Falls, WI 54022, USA}
\affiliation{Dept. of Physics and Astronomy, University of Rochester, Rochester, NY 14627, USA}
\affiliation{Department of Physics and Astronomy, University of Utah, Salt Lake City, UT 84112, USA}
\affiliation{Dept. of Physics, Chung-Ang University, Seoul 06974, Republic of Korea}
\affiliation{Oskar Klein Centre and Dept. of Physics, Stockholm University, SE-10691 Stockholm, Sweden}
\affiliation{Dept. of Physics and Astronomy, Stony Brook University, Stony Brook, NY 11794-3800, USA}
\affiliation{Dept. of Physics, Sungkyunkwan University, Suwon 16419, Republic of Korea}
\affiliation{Institute of Physics, Academia Sinica, Taipei, 11529, Taiwan}
\affiliation{Dept. of Physics and Astronomy, University of Alabama, Tuscaloosa, AL 35487, USA}
\affiliation{Dept. of Astronomy and Astrophysics, Pennsylvania State University, University Park, PA 16802, USA}
\affiliation{Dept. of Physics, Pennsylvania State University, University Park, PA 16802, USA}
\affiliation{Dept. of Physics and Astronomy, Uppsala University, Box 516, SE-75120 Uppsala, Sweden}
\affiliation{Dept. of Physics, University of Wuppertal, D-42119 Wuppertal, Germany}
\affiliation{Deutsches Elektronen-Synchrotron DESY, Platanenallee 6, D-15738 Zeuthen, Germany}

\author{R. Abbasi}
\affiliation{Department of Physics, Loyola University Chicago, Chicago, IL 60660, USA}
\author{M. Ackermann}
\affiliation{Deutsches Elektronen-Synchrotron DESY, Platanenallee 6, D-15738 Zeuthen, Germany}
\author{J. Adams}
\affiliation{Dept. of Physics and Astronomy, University of Canterbury, Private Bag 4800, Christchurch, New Zealand}
\author{S. K. Agarwalla}
\thanks{also at Institute of Physics, Sachivalaya Marg, Sainik School Post, Bhubaneswar 751005, India}
\affiliation{Dept. of Physics and Wisconsin IceCube Particle Astrophysics Center, University of Wisconsin{\textemdash}Madison, Madison, WI 53706, USA}
\author{J. A. Aguilar}
\affiliation{Universit{\'e} Libre de Bruxelles, Science Faculty CP230, B-1050 Brussels, Belgium}
\author{M. Ahlers}
\affiliation{Niels Bohr Institute, University of Copenhagen, DK-2100 Copenhagen, Denmark}
\author{J.M. Alameddine}
\affiliation{Dept. of Physics, TU Dortmund University, D-44221 Dortmund, Germany}
\author{S. Ali}
\affiliation{Dept. of Physics and Astronomy, University of Kansas, Lawrence, KS 66045, USA}
\author{N. M. Amin}
\affiliation{Bartol Research Institute and Dept. of Physics and Astronomy, University of Delaware, Newark, DE 19716, USA}
\author{K. Andeen}
\affiliation{Department of Physics, Marquette University, Milwaukee, WI 53201, USA}
% ...

\collaboration{IceCube Collaboration}
\email{analysis@icecube.wisc.edu}
\noaffiliation

\date{\today}

\begin{abstract}
The IceCube Upgrade is an extension of the existing IceCube Neutrino Observatory and will be deployed in the 2025-2026 austral summer. It will significantly improve the sensitivity of the detector to atmospheric neutrino oscillations. The existing 86-string IceCube array contains a dense in-fill known as DeepCore which is optimized to measure neutrinos with energies down to a few GeV. The IceCube Upgrade will consist of seven new densely-instrumented strings placed within the DeepCore volume to further enhance the performance in the GeV energy range. The additional strings will feature new optical modules, each containing multiple PMTs, in contrast to the existing modules that each contain a single PMT. This will more than triple the number of PMT channels with respect to the current IceCube configuration, allowing for improved detection efficiency and reconstruction performance at GeV energies. We describe necessary updates to simulation, event selection, and reconstruction to accommodate the higher data rates observed by the upgraded detector and the addition of multi-PMT modules. We determine the expected sensitivity of the IceCube Upgrade to the atmospheric neutrino oscillation parameters sin$^2\theta_{23}$ and $\Delta m^2_{32}$, the appearance of tau neutrinos and the neutrino mass ordering. The IceCube Upgrade will provide neutrino oscillation measurements that are of similar precision to those from accelerator experiments, while providing complementarity by probing higher energies and longer baselines, and with different sources of systematic uncertainties.
\end{abstract}


\maketitle

\section{Introduction}
\label{sec:intro}

\begin{figure*}[tb]
     \centering
     \vspace{-1em}
     \subfloat{
         \includegraphics[width=0.95\textwidth]{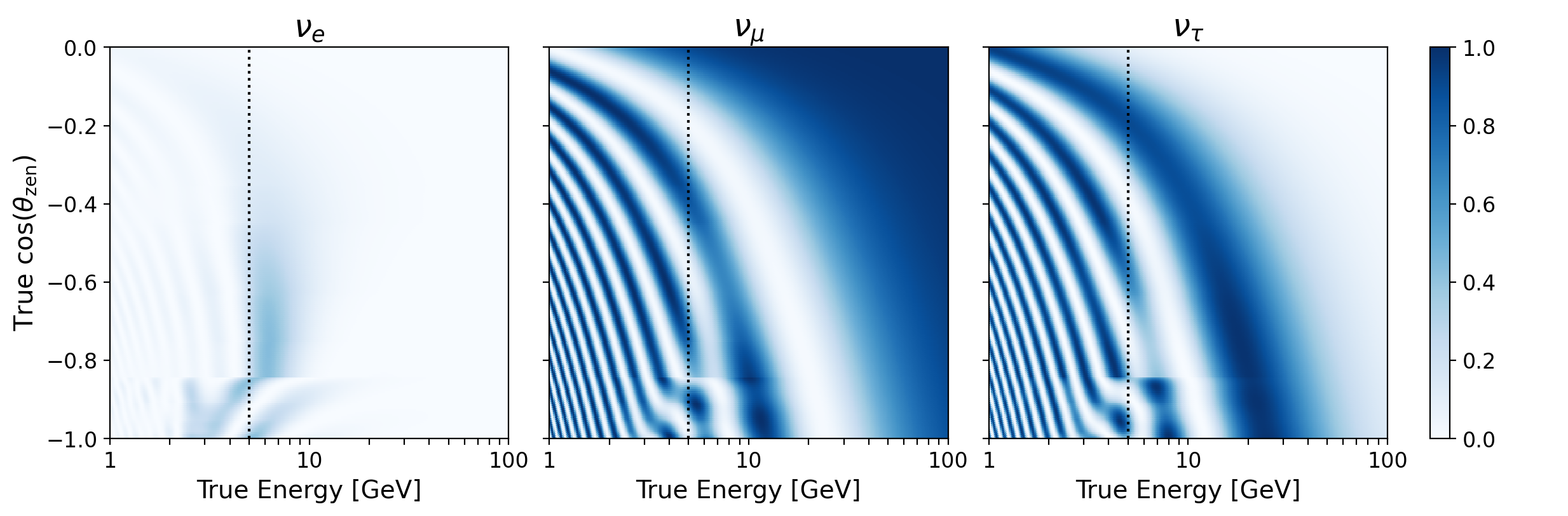}
         }
    \caption{Probability of a muon neutrino oscillating to each flavor (left: electron neutrino, center: muon neutrino, right: tau neutrino) as a function of true energy and arrival direction. cos($\theta_{zen}$) is a proxy for the distance that the neutrino traveled. The IceCube Upgrade will make the region below 5 GeV, indicated by the dotted line, much more accessible to analyses.} 
    \label{fig:osc_probs}
\end{figure*}

Studying the oscillations of atmospheric neutrinos provides a unique window into the nature of these elusive particles. Since neutrinos arrive at the detector from all directions, many different oscillation baselines are probed simultaneously ranging from tens of kilometers up to the diameter of the Earth. This can be used to make precise measurements of the atmospheric neutrino oscillation parameters, such as the previous measurements in~\cite{IceCube:2024oscnext,SuperK}, as described by the Pontecorvo-Maki-Nakagawa-Sakata (PMNS) mixing matrix \cite{Pontecorvo:1957cp,Maki:1962mu}. Figure~\ref{fig:osc_probs} shows the probability of an atmospheric muon neutrino oscillating into each flavor as a function of energy and zenith angle, where zenith angle is a proxy for distance traveled. The primary signature is the disappearance of muon neutrinos and the appearance of tau neutrinos. There is negligible production of tau neutrinos in the atmosphere, and thus detected tau neutrinos in this energy region arise from oscillations. The primary physics parameters $\Delta m^2_{32}$ and $\theta_{23}$ control the location and amplitude of the peaks, respectively. The distortions visible below cos$(\theta_{zen})<-0.8$ corresponds to the region where neutrinos pass through the core of the Earth, which has much higher density. These matter effects are a sub-dominant component that occur in all flavors primarily below 10 GeV, however they are most noticeable in the electron neutrino channel since the other channels are dominated by the $\nu_\mu\rightarrow\nu_\tau$ oscillation. In addition, the different effects of matter on neutrinos and anti-neutrinos traversing the Earth can be used to distinguish the ordering of the neutrino mass eigenstates \cite{Blennow:2013oma}. Therefore, the neutrinos in this region below 10 GeV will be crucial for measurements of the neutrino mass ordering.

Atmospheric neutrinos are generated in cosmic-ray interactions with nucleons in the Earth's atmosphere. When interacting in optically clear media, such as water or ice, the interactions of these GeV neutrinos produce charged secondary particles that emit Cherenkov radiation. The emitted photons can be seen by ice- or water-Cherenkov detectors such as the IceCube Neutrino Observatory \cite{Aartsen:2016icecube} which is located at the geographic South Pole. IceCube currently consists of 5,160 photosensors, known as Digital Optical Modules (DOMs), spread across 86 long cables, known as strings, each deployed in a separate borehole in the ice.

The primary IceCube array is spread throughout a cubic-kilometer volume and was designed to observe high-energy neutrino interactions at the TeV to PeV energy scale. These strings are arranged in a hexagonal grid. The existing low-energy extension, known as IceCube DeepCore \cite{IceCube:2011deepcore}, consists of eight strings with more dense optical module spacing and high quantum efficiency photomultiplier tubes (PMTs). The DeepCore modules are located in the deepest and clearest portion of the ice. The optical module spacing in DeepCore allows for the detection of neutrinos between roughly 5\;GeV to 500\;GeV at a rate of $\mathcal{O}$(mHz) at analysis level. DeepCore has already collected about a decade of atmospheric neutrino data, with recent neutrino oscillation measurements reported in~\cite{oscnext_prd,IceCube:2024oscnext}.

To lower the energy threshold and increase the rate of detected neutrinos, seven additional strings will be deployed within the DeepCore fiducial volume in the 2025-2026 austral summer. The installation of these additional strings, known as the IceCube Upgrade, will increase the total number of strings in IceCube from 86 to 93. This fully funded IceCube Upgrade extension is distinct from the previously proposed ``PINGU'' extension (Precision IceCube Next Generation Upgrade) \cite{IceCube-PINGU:2014} which is no longer being pursued.

The additional strings will be equipped with new optical modules, each hosting multiple PMTs, as well as several new calibration devices. After installation, the number of PMT channels will be more than three times higher than the current detector configuration. Additionally, the total photocathode area in the DeepCore fiducial volume will increase by a factor of about three. The effective photosensitivity increase will be even higher than this due to improvements in module design that increase the acceptance for shorter wavelengths, discussed in more detail in Section \ref{sec:optical_modules}. These improvements together with closer module spacing will greatly enhance IceCube's capabilities in the GeV energy regime.

This paper discusses the IceCube Upgrade detector and simulation (Section~\ref{sec:icu}), new tools developed for handling IceCube Upgrade events (Sections~\ref{sec:simulation} and \ref{sec:data_sample}), and shows the expected enhancement in sensitivity for several benchmark oscillation analyses by comparing a scenario with and without the additional strings (Sections~\ref{sec:dc_icu} and \ref{sec:sensitivities}).

\section{The IceCube Upgrade Detector}
\label{sec:icu}

\subsection{Geometry}

\begin{figure*}
    \centering
    \includegraphics[width=\textwidth]{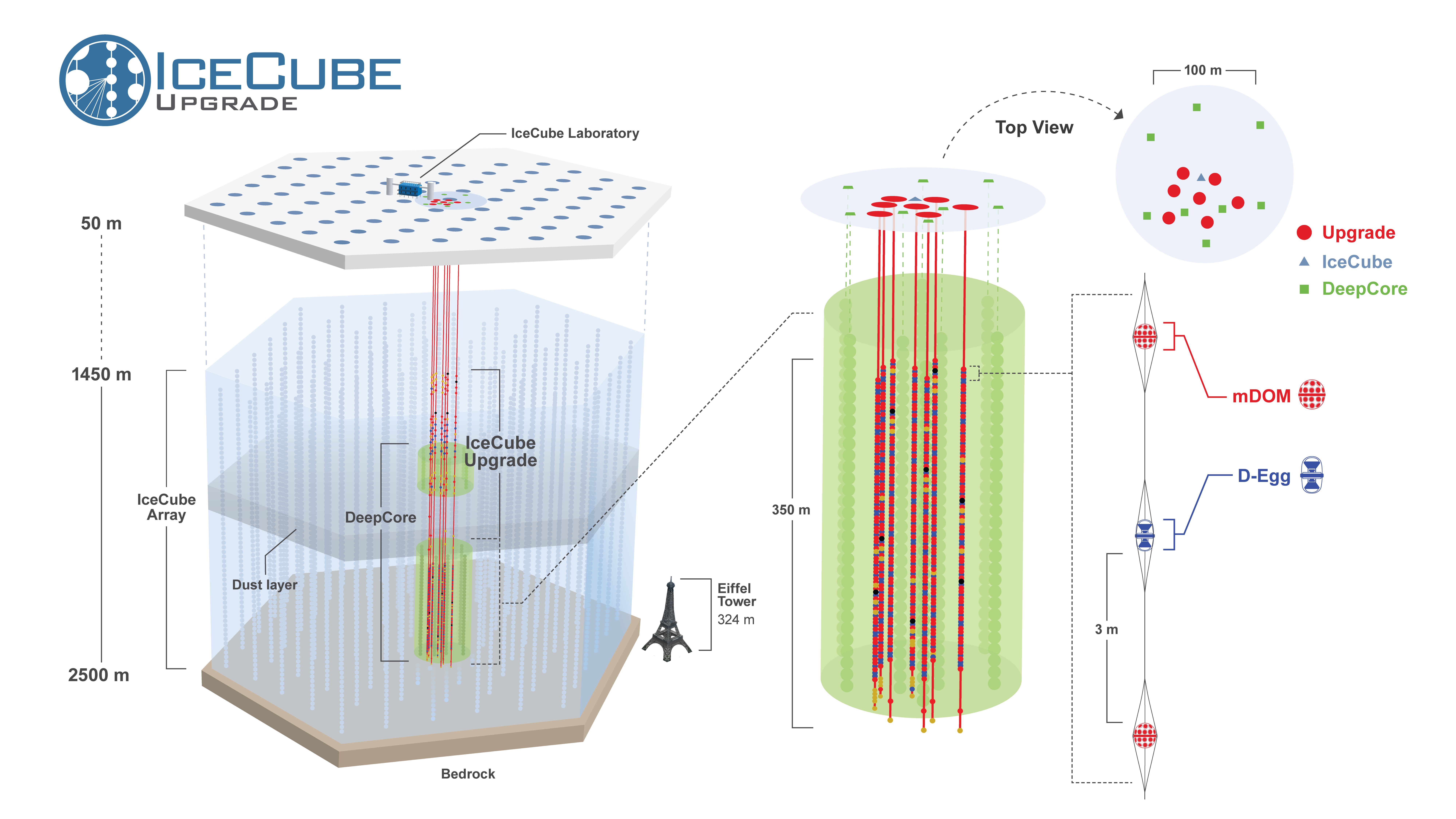}
    \caption{The IceCube Upgrade will consist of 7 new strings, to be deployed in 2025-2026, at the locations shown in the top view in red. The current IceCube array optimized for TeV neutrinos is shown in blue, while the existing low-energy extension known as DeepCore is shown in green. 
    The side view shows the distribution of module types on the new strings. The main optical modules listed in Table~\ref{table:detector_summary} are shown. The modules shown in yellow and black correspond to the R\&D optical modules described at the end of Section~\ref{sec:optical_modules} and the calibration modules described in Section~\ref{sec:calibration_modules}. Most of the modules in the IceCube Upgrade are located in the deepest part of the ice, but there are some modules in the upper portion of IceCube that will primary be used for calibration.}
    \label{fig:geometry}
\end{figure*}

The IceCube Upgrade will consist of seven new strings deployed within the DeepCore fiducial volume, as shown in Fig.~\ref{fig:geometry}. The existing spacing between strings in the DeepCore fiducial volume is about 40-70\;m. With the IceCube Upgrade, the typical inter-string spacing in this region will be reduced to about 20-30\;m. In addition, the new modules will be more closely spaced in depth. On the IceCube strings, the modules are spaced 17\;m apart vertically spanning depths of about 1450\;m to 2450\;m below the surface. The modules on the DeepCore strings are spaced 7\;m apart and are concentrated in the deepest part of the ice from 2100\;m to 2450\;m, where the ice has the best optical properties. The modules on the Upgrade strings will be spaced 3\;m apart, with most of the modules concentrated between 2160\;m and 2430\;m.

To improve our understanding of the optical properties of the ice, there will also be modules between 1300\;m and 1900\;m as well as below 2450\;m which will be used for calibration purposes. The existing IceCube plus DeepCore configuration is referred to as ``IC86''. The configuration including the new IceCube Upgrade strings is referred to as ``IC93''.

\subsection{Optical modules}
\label{sec:optical_modules}

\begin{table*}[hbt!]
  \centering
  \caption{Key parameters of the two main Upgrade detectors D-Egg and mDOM compared with the existing IceCube DOM}
  \label{table:detector_summary}
    \begin{ruledtabular}
    \begin{tabular}{ccccccccc}
    name & \# modules & diameter & PMT size & \# PMTs & \# flasher LEDs & \# cameras\\
    \hline 
    D-Egg & 280 & 300 mm & 8 inch & 2 & 12 & 3 \\
    mDOM & 400 & 356 mm & 3 inch & 24 & 10 & 3 \\
    IceCube DOM & 5160 & 350 mm & 10 inch & 1 & 12 & -\\ 
    \end{tabular}
    \end{ruledtabular}
\end{table*}

\begin{figure}[tb]
    \centering
    \subfloat{
        \includegraphics[width=0.14\textwidth]{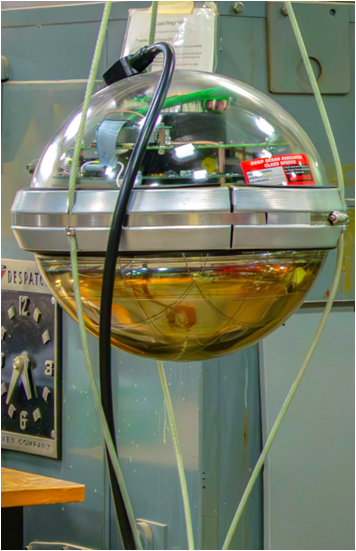}
        }
    \subfloat{
        \includegraphics[width=0.14\textwidth]{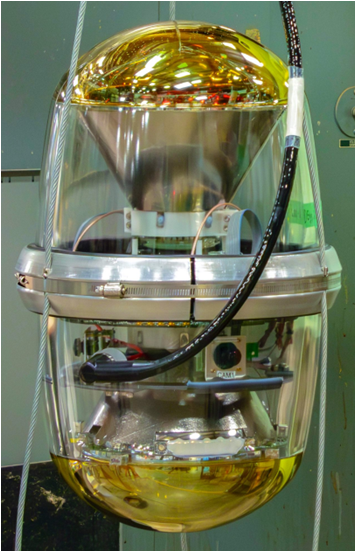}
        }
    \subfloat{
        \includegraphics[width=0.14\textwidth]{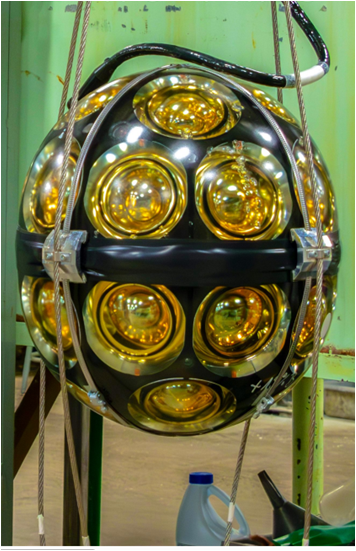}
        }
    \caption{Photographs of the modules: IceCube DOM (left), D-Egg (middle) and mDOM (right).}
    \label{fig:modules}
\end{figure}

The IceCube Upgrade strings will be equipped with new types of optical modules. Two different sensor types form the majority of the approx. 700 new optical modules that will be deployed. There will be 280 ``D-Eggs'' (Dual optical sensors in an Ellipsoid Glass for Gen2) \cite{IceCube:2022degg} each containing two 8-inch R5912-100-70 PMTs, and 400 ``mDOMs'' (multi-PMT Digital Optical Module) \cite{mdom_icrc2019,mdom_icrc2021,mdom_icrc2023} each containing 24 3-inch R15458-02 PMTs \cite{IceCube:2024mdom_pmts}. The distribution of the new modules across the detector is shown in Fig.~\ref{fig:geometry}. Photographs of the modules are shown in Fig.~\ref{fig:modules}.

Both modules have several advantages over the existing IceCube and DeepCore modules with only one 10-inch PMT each. The multiple-PMT configuration provides for a more uniform solid angle coverage, increases the photocathode area per module, and offers an intrinsic directional resolution. D-Egg and mDOM PMTs were selected to have low dark noise rates, high quantum efficiency, and excellent charge/timing resolution. Key parameters of these modules are summarized in Table \ref{table:detector_summary}. 

D-Eggs and mDOMs have fast, low-power electronics that digitize PMT analog signals at 240 MHz and 120 MHz, respectively, to achieve nanosecond precision. The readout system has no dead time using internal 2GB DDR3 RAM for continuous data acquisition. The high noise rates from multiple PMTs, however, would saturate the communication bandwidth of each string's cable. Microcontrollers inside the modules partially mitigate this effect through lossless compression, feature extraction of single photon events, and in-module local coincidence.

Both module types incorporate several calibration devices. Each module has LEDs that produce nanosecond pulses pointing at various orientations which will be used for PMT gain calibration and ice property inspection using neighboring modules \cite{IceCube:2025ylq_LEDs}. They are also equipped with fisheye cameras and illumination LEDs to monitor the refrozen ice surrounding the strings \cite{Kang:2023kjl}, as well as environmental sensors to monitor pressure, temperature, magnetic field, and acceleration.

Each module's components are housed in a pressure-tolerant glass vessel. The PMTs are optically coupled to the vessels using silicone gel for increased photon transparency and mechanical support. The vessels and the gels were selected based on their UV transparency, greatly enhancing the transmittance at shorter wavelengths compared to the existing modules, as well as their low radioactivity. For the D-Egg, the overall high UV transparency, together with the increased photocathode area and the higher quantum efficiency of the PMTs, leads to the Cherenkov-averaged photon sensitivity about 2.8 times higher than the existing IceCube DOMs \cite{IceCube:2022degg}. The D-Egg and mDOM glass vessels have diameters of 300\;mm and 356\;mm, respectively, comparable to or smaller than the existing 350\;mm DOMs.

In addition to D-Eggs and mDOMs, some other types of optical modules will be deployed for R\&D purposes. These modules include pDOMs (DeepCore-like single PMT DOMs), WOMs (Wavelength shifting Optical Modules), LOMs (Long Optical Modules), FOMs (Fiber Optic Module). pDOMs each contain one high quantum efficiency PMT and a muon-tagging assembly. WOMs \cite{Bastian-Querner:2021uqv} use wavelength shifting paint in a transparent tube to increase the photon detection rate at short wavelengths. LOMs \cite{IceCube-Gen2:2023rji,IceCube-Gen2:2023ycx} are a hybrid between the D-Egg and mDOM each containing 16 or 18 4-inch PMTs with a narrower diameter than the existing IceCube modules. The FOM uses florescent optical fibers to guide photons to a PMT. The R\&D modules are not included in the sensitivity calculations that follow.

\subsection{Calibration modules}
\label{sec:calibration_modules}

The IceCube Upgrade strings will also feature a suite of about 50 new calibration devices aimed at improving the characterization of our detection medium, the glacial Antarctic ice. The Precision Optical Calibration Modules, or ``POCAMs'' \cite{pocam,pocam_upgrade}, are isotropic light sources providing nanosecond light pulses that can be used for optical calibration. ``Pencil Beam'' modules \cite{Rongen:2021rgc} produce a laser-like beam that can be pointed in almost any direction, allowing for more precise directional and anisotropy calibrations. Acoustic modules \cite{IceCube:2021zse} will be used for the precise calibration of the positions of the optical sensors, and camera systems \cite{Kang:2023kjl} will observe the optical properties of the ice. In addition, seismometers \cite{seismometer} allow for a broad spectrum of Earth science studies.

\section{Simulation}
\label{sec:simulation}

\subsection{Particle Interaction and Propagation}
\label{sec:particle}

The tools used to produce Monte Carlo (MC) simulation for the IceCube Upgrade are similar to those used for previous IceCube DeepCore measurements \cite{oscnext_prd}. The simulation chain has been improved since the previous IceCube Upgrade public data release \cite{upgrade_data_release_2020}. It now includes a more realistic detector response and improved reconstruction and processing.

The simulation chain consists of three steps: primary particle interaction, photon (and other particle) propagation, and detector response. We simulate three different types of events: signal neutrino events and backgrounds from atmospheric muon events and pure noise events. The latter arise from random coincidences of PMT noise.

Neutrino events are generated using \textsc{genie} version 2.12.8 \cite{genie:2009,genie_v2} following an $E^{-2.5}$ energy spectrum. They are produced throughout a cylindrical volume surrounding the DeepCore region, which also includes the new Upgrade strings. We simulate all three neutrino flavors, neutrinos and anti-neutrinos, charged current (CC) and neutral current (NC) interactions. Neutrino events are weighted to a nominal atmospheric neutrino flux from Honda \textit{et al.} \cite{Honda:2015fha} plus a three-flavor oscillation probability including matter effects.

Atmospheric muons are simulated using \textsc{MuonGun} (based on \cite{Becherini:2005sr}) which produces muons on the surface of a cylinder surrounding IceCube and then weighted to \textsc{corsika} \cite{Heck:1998vt}. Computation time is reduced by requiring that the muon reaches an inner cylinder surrounding the DeepCore/Upgrade region. The energy and zenith angle (the angle between the muon direction and the Earth's axis) of the generated muons are biased to prioritize muons that are most likely to survive the later stages of the event selection, following the procedure introduced in \cite{kayla_thesis}. The muon events are reweighted accordingly to avoid bias in the final event distributions.

After the primary particle interaction, muons from $\nu_\mu$ CC interactions and atmospheric muons are propagated using \textsc{proposal} \cite{proposal}. All other particle types are propagated and decayed using \textsc{Geant4} \cite{GEANT4:2002zbu}. The charged particles from these interactions with sufficiently high energy will emit Cherenkov radiation. The resulting Cherenkov photons are then propagated through the ice using \textsc{CLSim} \cite{clsim} according to the ice model described in \cite{paper_bfr}.

Muons travel about 4.5\;m per GeV in ice and therefore muons detected by the IceCube Upgrade can travel several tens of meters. In contrast, other leptons and hadrons travel much shorter distances. As a result, the distribution of light in the detector looks different depending on whether a muon was present or not. Consequently, $\nu_\mu$ CC interactions as well as atmospheric muons produce an elongated event topology which we refer to as ``tracks'', while all other interaction types will produce rather spherical events, denoted ``cascades''. Neutrinos and anti-neutrinos are not distinguished in the detector on an event-by-event basis; only the sum of both contributions is observed.

\subsection{Detector Response}
\label{sec:detectorresponse}

Once a simulated photon reaches the surface of a new module, the following detector response is applied. First, the photon detection probability is determined based on the module angular acceptance \cite{holeice_icrc2023} and quantum efficiency. If the photon is detected, it is converted into a photoelectron (PE). In addition to these signal PEs, noise PEs from PMT thermal dark noise and radioactive decays in the glass pressure housing and PMT glass are added. PEs that are closer in time than 0.2 ns are merged since these would not be resolved separately by readout electronics. Then, the PEs are given to a PMT response simulation which outputs ``pulses" that have a charge and a time value assigned. The charge is randomly drawn from a Gaussian+exponential PMT charge distribution \cite{Aartsen:2016icecube}, while the time is the first PE time smeared by the PMT transit time spread. Pulses with a charge of less than 0.25 of the average single PE charge are removed because such pulses would not be distinguishable from noise in the electronics. Finally, a Gaussian smearing with a width of 1\;ns is applied to the time of each pulse to account for the readout timing resolution. A second Gaussian smearing with a width of 1\;ns is applied to all pulses in a module simultaneously to emulate the synchronization timing resolution.

In contrast to this simplified detector response, for the existing IceCube DOMs the actual waveform is simulated and unfolded to extract pulse information (see Section 4 in \cite{icecube_wavedeform}). The detector response of the new modules will be updated to reach the same level of detail in the future.

Random pure noise triggers are simulated as well. Since they do not involve a primary particle nor a photon propagation step, pure noise events only go through the detector response step and only involve noise PEs.

\section{Simulated Data Pipeline}
\label{sec:data_sample}

\subsection{Event Trigger}
\label{sec:trigger}

Once the pulses are recorded, a trigger is used to look for spatial and temporal coincidences to determine whether an event occurred, and if so, which pulses belong to it. Two different triggers are used and events are kept if either one or both of them fired. One is the DeepCore simple majority trigger ``SMT3'' described in \cite{IceCube:2011deepcore} which only uses existing IceCube and DeepCore strings, while the second uses the new Upgrade strings and is described in the following.

 Pulses that are near each other in space and time are used to establish ``local coincidence'' (LC), indicating that they could be causally connected. Different coincidence criteria are used for checking multiple PMTs on the same optical module and for multiple optical modules on the same string. For the first case, a pair of pulses are defined as coincident if they are on same module but on different PMTs and their time difference is less than 10\;ns. Two pulses are also coincident if the modules are close to each other on the string, defined as $\pm$ 8 modules ($\sim$24\;m), and if their time difference is less than 250\;ns. This corresponds approximately to the speed of light in ice. The pulses with ``local coincidence'' are then used to check for triggers. If 8 or more LC pulses occur within a 1750\;ns window, the trigger fires. All pulses from 4\;$\mu$s before to 6\;$\mu$s after the trigger condition is met are read out. The dominant event types triggering the detector are atmospheric muons $O$(100Hz) and pure noise events $O$(10Hz). Further levels of processing are necessary to reduce these backgrounds, and are described in Section \ref{sec:event_selection}.

\subsection{Removal of Noise Pulses}
\label{sec:noise_cleaning}
\label{sec:clean}

In the past, IceCube has been relying on cuts on the spatial and temporal distance between pulses to remove noise pulses from events. However, due to the increased number of PMTs in the new modules and higher than anticipated level of radioactivity in the mDOM PMT glass, these cut-based methods have been insufficient for noise removal.
To address this, we have developed a new “noise cleaning” algorithm based on Graph Convolutional Neural Networks (GNNs). GNNs work on \textit{graphs} and can be thought of as generalized convolutional neural networks. A graph is a collection of nodes and edges, where nodes typically represent data points and edges denote a relationship between the nodes, defining how information may spread during convolutions \cite{gnn_review}.

\begin{table}[bt]
  \centering
  \caption{\label{table:input_features} Input features used for training pulse cleaning and reconstruction methods.}
    \begin{ruledtabular}
    \begin{tabular}{ccc} 
    Feature & Description & Dimension \\
    \hline 
    $P_{xyz}$   & PMT position  & $\mathbb{R}^3$\\
    $D_{xyz}$  & PMT orientation unit vector & $\mathbb{R}^3$ \\
    $t$ & Estimated arrival time at PMT  & $\mathbb{R}$ \\
    $q$  & Estimated PE charge & $\mathbb{R}$ \\ 
    $rde$ & Relative DOM Eff.  &  $\mathbb{R}$ \\
    $A$ & Area of PMT  & $\mathbb{R}$  \\
    $idx$ & String, DOM and PMT number  & $\mathbb{Z}^3$  \\
    $type$ & DOM type  & $\mathbb{Z}$  \\
    \end{tabular}
    \end{ruledtabular}
\end{table}

We chose to represent IceCube events as point cloud graphs, where nodes represent each PMT pulse recorded in an event and edges are drawn to the 8 nearest neighbors of each node based on the Euclidean distance between PMTs. The features associated with each node are shown in Table \ref{table:input_features}. Using this graph representation, we then perform binary classification to classify each node (pulse) in each event as either signal or noise. We train a modified version of DynEdge \cite{icecube_dynedge} from the open-source deep learning library for neutrino telescopes GraphNeT \cite{graphnet_paper}. This new noise cleaning method produces a prediction between 0 and 1 for each pulse in an event, where 0 indicates a pulse is likely noise and 1 indicates a pulse is likely signal. The method is compatible with events containing any number of pulses.

\begin{figure}[tb]
    \centering
    \includegraphics[width=0.5\textwidth]{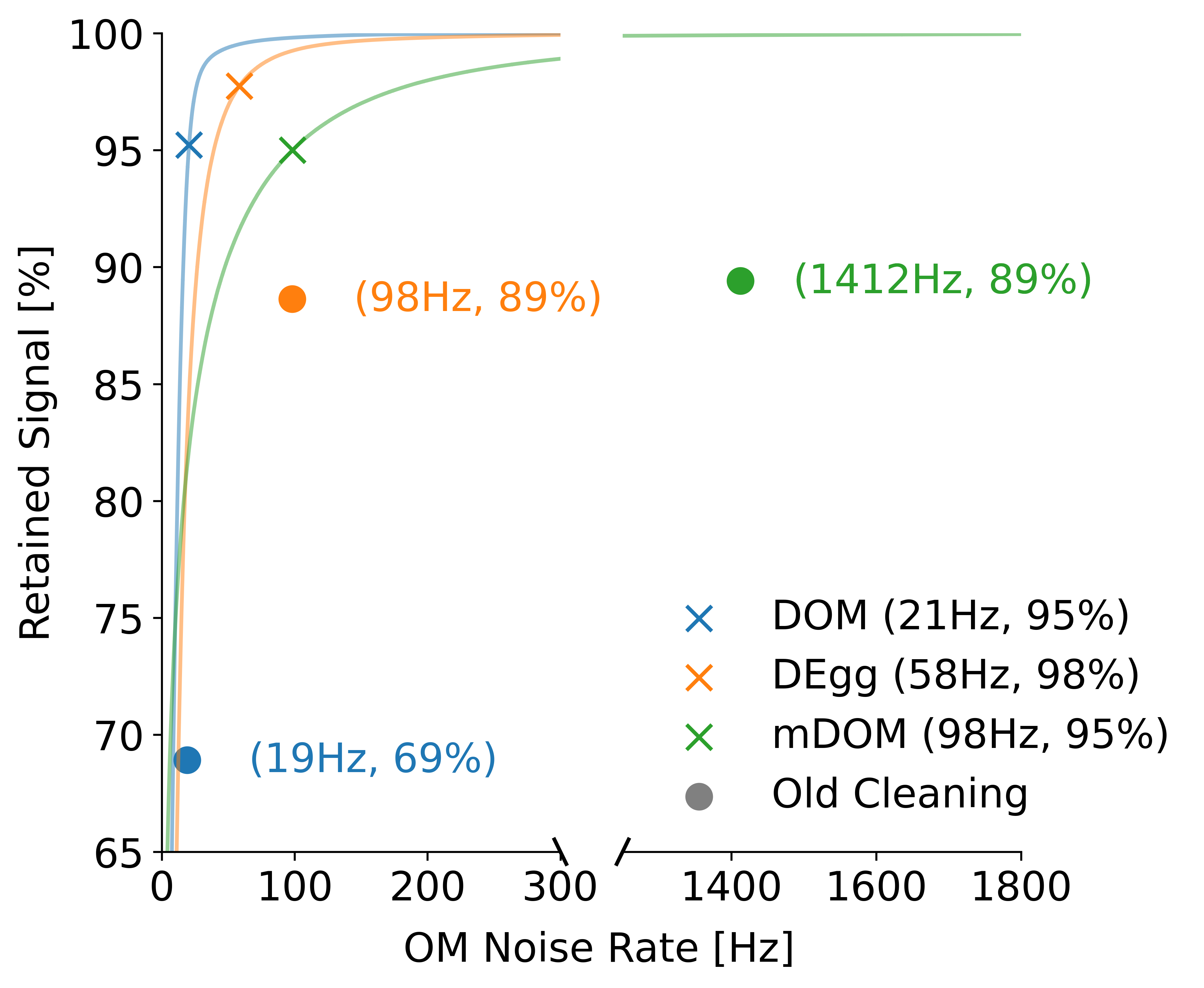}
    \caption{Retained signal vs. module noise rate. The retained signal is the true positive rate in percent, and the optical module noise rate is computed as the false positive rate multiplied by the total optical module noise rate. Each curve represents different optical module types. Crosses denote the classification threshold of 0.7 used for this analysis. The performance of the previous cleaning method for each module type is added as a circular dot in the respective color for reference.}
    \label{fig:noise}
\end{figure}

Because the output score is continuous, the amount of cleaning can be easily tuned by changing the threshold. Once trained on an independent sample of close to 4 million neutrino and muon events in IceCube Upgrade, we find that, at a threshold of 0.7, the GNN on average reduces the amount of noise by roughly a factor 10 with only minor loss ($\sim$5\%) in signal hits. The contamination of noise for pulses that pass the cleaning is 6.8\% using the new GNN method, whereas for the traditional noise cleaning method applied to the Upgrade it was 70\%. The previous method looks for hits that could be causally connected and is described in \cite{oscnext_prd}. The retained signal vs. optical module noise rate is shown for the GNN-based cleaning method and compared to the traditional cleaning method in Fig.~\ref{fig:noise}.  As seen in Fig.~\ref{fig:noise}, the GNN cleaning method retains a higher percentage of the signal than the traditional method, and the remaining noise rate depends on the module type. The pulses that remain after cleaning are referred to as the ``cleaned pulse series''. 

\subsection{Event Reconstruction}
\label{sec:event_reconstruction}
 Four reconstruction and classification tasks are trained using different instances of DynEdge with a procedure similar to that outlined in \cite{icecube_dynedge}. In order to mitigate class imbalances during training, task-specific sub-samples produced from the full simulation sample are used for training. These samples differ primarily in their composition of neutrino track ($\mathcal{T}$) events, neutrino cascade ($\mathcal{C}$) events, and muon ($\mu$) background events, as seen in Table \ref{table:training_samples}. The \textit{Reco} sample contains neutrino events with at least three pulses in the cleaned pulse series. The $\nu/\mu$ sample was created by first subsampling the whole sample to obtain an equal number of neutrino and muon events and subsequently requiring three pulses in the cleaned pulse series. An identical procedure was used to construct the $\mathcal{T}/\mathcal{C}$ sample, but $\nu_\tau$ CC events have been omitted, as around 17\% of these interactions may produce track-like signatures, which may affect training \cite{pdg, icecube_dynedge}.  
\begin{table}[hbt!]
  \centering
  \caption{Overview of task-specific subsamples used for training. Samples marked with * are uncleaned events, whereas all other samples have been cleaned by the GNN-based method. ``Size'' here indicates the number of simulated events in the sample.}
  \label{table:training_samples}
    \begin{ruledtabular}
    \begin{tabular}{ccc} 
    Task & Size (millions) & Composition \\
    \hline 
    Pulse Cleaning   &  4.04*  & $\mathcal{T} (41\%)$, $\mathcal{C} (51\%)$, $\mu (8\%)$ \\
    $\nu / \mu$  &  0.367  &  $\mathcal{T} (27\%)$, $\mathcal{C} (32\%)$, $\mu (41\%)$\\
    $\mathcal{T} / \mathcal{C}$  & 1.81 & $\mathcal{T} (53\%)$, $\mathcal{C} (47\%)$ \\
    Reco  & 2.39 & $\mathcal{T} (46\%)$, $\mathcal{C} (54\%)$ \\ 
    \end{tabular}
    \end{ruledtabular}
\end{table}

\begin{figure*}[ht]
     \centering
     \vspace{-1em}
     \hspace*{-1cm}
     \subfloat{
         \includegraphics[width=0.28\textwidth]{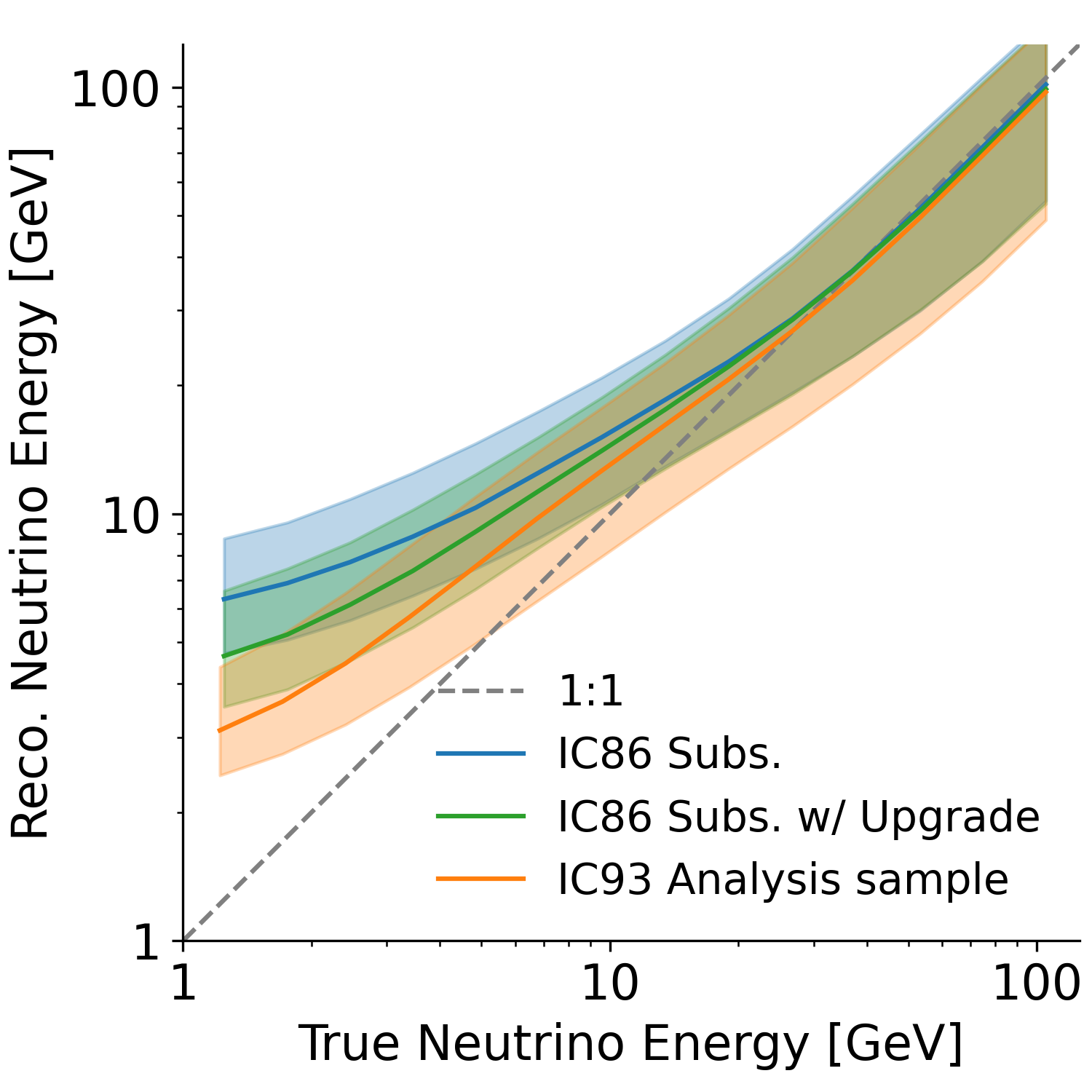}
    }
     \hspace*{1cm}
     \subfloat{
         \includegraphics[width=0.28\textwidth]{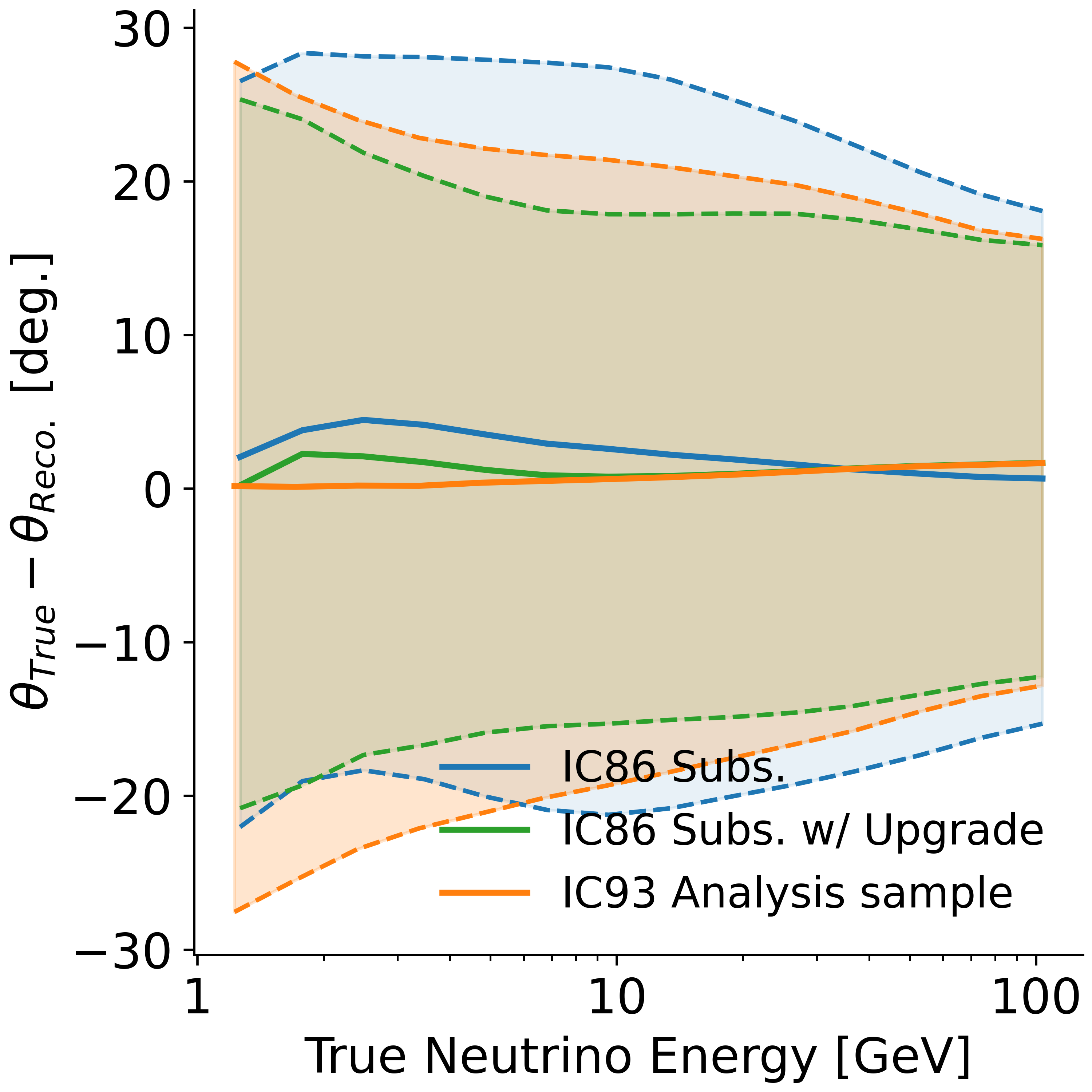}
     }
     \hspace{1cm}
     \subfloat{
         \includegraphics[width=0.28\textwidth]{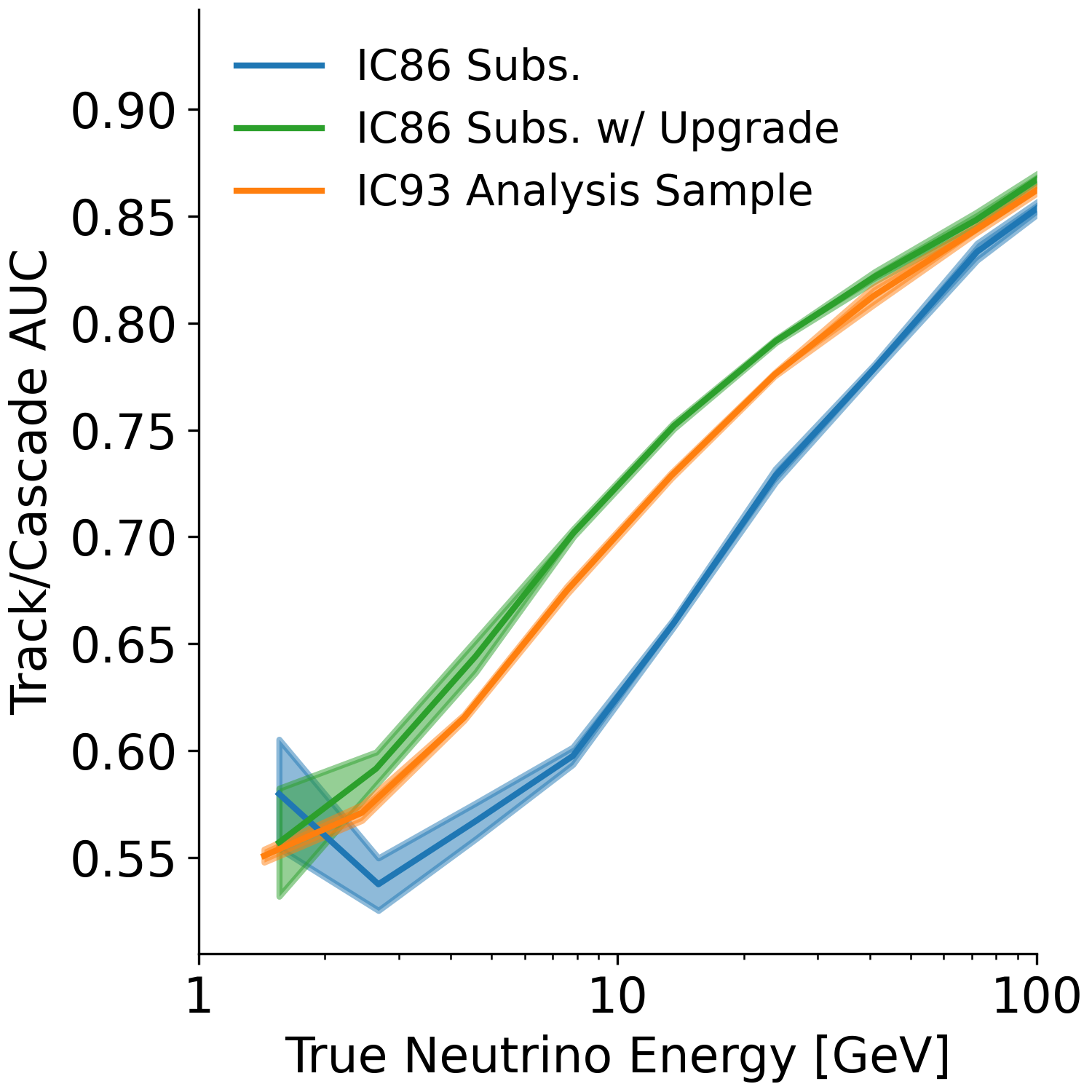}
     }

        \caption{Performance of reconstruction models on analysis sample (orange) and the IC86 subsample with and without the information on the Upgrade strings (green and blue, respectively). Left: Reconstructed neutrino energy vs. true neutrino energy. Bands depict the 1$\sigma$ range, and the median is shown in solid. Middle: Angular resolution of the zenith reconstruction. Bands represent the  1$\sigma$ range of the residual distribution around the median shown in solid. Right: Area under curve (AUC) score of $\mathcal{T}$/$\mathcal{C}$ classifier binned in neutrino energy. Uncertainty bands on the AUC curves result from bootstrapping and represent the expected statistical 1 $\sigma$ deviation. A value of 0.5 indicates random classification, and a value of 1.0 indicates perfect classification. There is a moderate improvement in energy and zenith reconstruction, while the improvement in $\mathcal{T}$/$\mathcal{C}$ classification is quite substantial especially around 10 GeV.}
        \label{fig:reco}
\end{figure*}

In contrast to the noise cleaning task, where each pulse is assigned an output score, event classification and reconstruction are run on a per-event mode, where each model produces a single prediction for each event. The first model is trained to estimate neutrino energy and uses the $Reco$ sample mentioned in Table~\ref{table:training_samples}. Another model is trained to estimate the zenith angle of the incoming neutrino, along with an uncertainty estimation on the zenith prediction. The zenith reconstruction network is also trained using the $Reco$ sample. 

The remaining tasks are classification. For the third event-level task an instance of DynEdge is trained to classify events as track-like or cascade-like using the $\mathcal{T} / \mathcal{C}$ sample. The fourth event-level task is classifying between muon and neutrino events, which is used in the event selection to reject background muons from the neutrino sample. This is trained on the  $\nu / \mu$ sample. To mitigate class imbalance, the  $\nu / \mu$ sample has a roughly equal number of muon and neutrino events in it. Due to the computational intensity of producing background muon simulation, we have a relatively low number of simulated muons, and therefore this sample is the smallest at just 367k events.

Selected performance metrics and their energy dependence are shown in Fig.~\ref{fig:reco} for the zenith and energy reconstruction models, along with the energy dependence of the  $\mathcal{T} / \mathcal{C}$ classifier's performance. Fig.~\ref{fig:reco} compares the performance metrics between IC86 and IC93. However, because the threshold for detection by IC86 is higher than IC93, the IC93 sample (orange) includes low-energy events that were not detectable by IC86; these low-energy events tend to produce only a few pulses in the detector and are, therefore, hardest to reconstruct. In order to directly quantify the improvement due to the new strings on the exact same set of events, we identify a subset of events detectable by IC86 that have at least one pulse in IC93. The events in this sub-sample detectable by IC86 are then reconstructed with and without the information on the new strings, denoted ``IC86 Subs. w/ Upgrade" (green) and ``IC86 Subs." (blue) respectively. Comparing the green and blue thus represents the improvement on the exact same events directly due to the inclusion of the new strings. Additional details about these reconstruction comparisons can be found in Appendix~\ref{sec:appendix_reco}.

\subsection{Event Selection}
\label{sec:event_selection}
\begin{figure}[htb]
    \centering
    \includegraphics[width=0.44\textwidth]{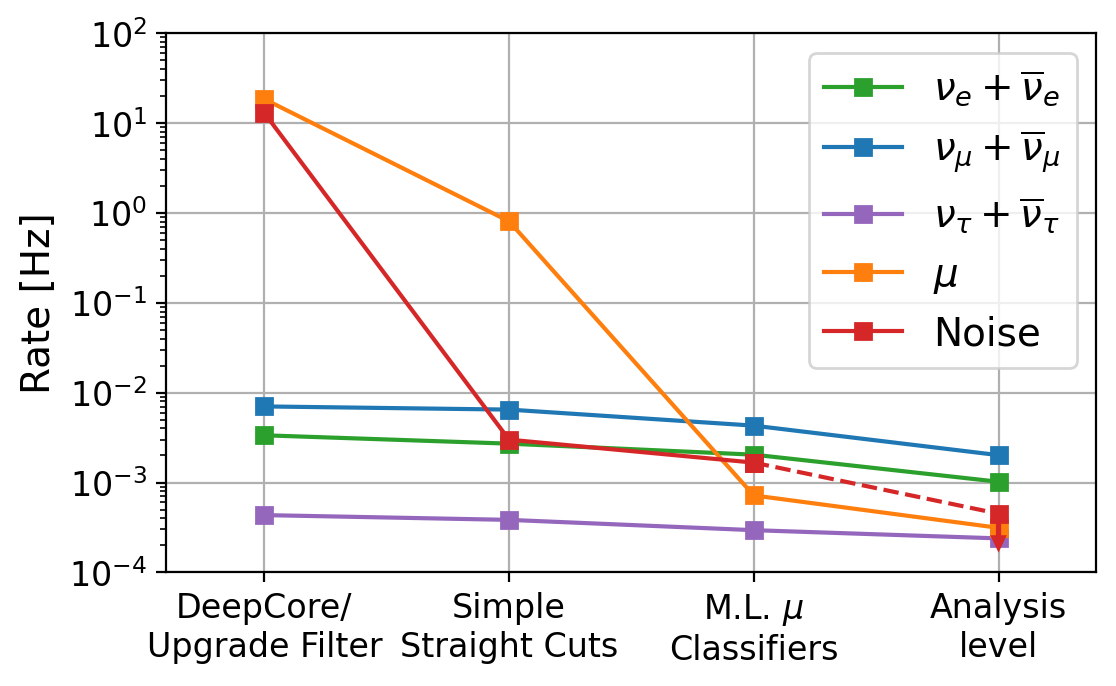}
    \caption{Signal and background event rates for the different levels of selection. For the noise rate at analysis level, the 90\% upper limit is shown (indicated by the dashed line and the arrow). NuFit5.2 \cite{nufit_5_2} and the normal ordering are used to calculate the neutrino rates.}
    \label{fig:rates}
\end{figure}

At trigger level, the events are dominated by atmospheric muons and pure noise triggers, by several orders of magnitude compared to atmospheric neutrino rates. Further levels of event selection are therefore required to obtain a neutrino-dominated event sample. These levels as well as their rates are summarized in Fig.~\ref{fig:rates}.

The IC93 filter is adapted from the DeepCore filter~\cite{oscnext_prd} in order to include the new strings. The filter uses a veto region around DeepCore to eliminate muon events which leave light in this region. For each triggered event, the pulses are split into those occurring in the fiducial volume and the veto region. For pulses in the fiducial region, the center of gravity in space and time is calculated. If one of the pulses in the veto region is causally connected to this center of gravity, the event is rejected.

Next, a series of simple straight cuts eliminate the most obvious remaining background events. Cuts on five different variables are used. Four of them are based on the cleaned pulse series from the noise cleaning procedure described in Section~\ref{sec:noise_cleaning}: the number of modules with pulses, the number of pulses in the fiducial volume, the z-position of the first module that saw light, and the time difference between the first and the last pulse. The fifth variable is the time difference between the first and the last pulse before noise cleaning. These variables were selected based on previous experience with DeepCore samples and were also used in \cite{oscnext_prd} with similar cut values. The pure noise events are already suppressed by these cuts by more than three orders of magnitude, mainly by the first two cuts.

Next, a collection of machine learning classifiers is used to identify muon background events. One classifier is the GNN described in Section~\ref{sec:event_reconstruction}, the other two are boosted decision trees (BDTs) that use reconstructed quantities and variables describing the spatial and temporal distribution of the pulses. More details about the BDTs can be found in \cite{jorge_thesis}. Applying all three classifiers leads to a reduction of the remaining muon background by three orders of magnitude.
Table~\ref{table:event_rates} compares the IC93 neutrino rates that we obtain through this selection with the rates from IC86 which use a selection similar to \cite{IceCube:2024oscnext} and \cite{oscnext_prd}. The IC93 rates are significantly increased with respect to IC86.

\begin{table}[h]
  \centering
  \caption{Neutrino rates after event selection (but before final analysis level cuts) for the current (IC86) and upgraded (IC93) detector setting.}
  \label{table:event_rates}
    \begin{ruledtabular}
    \begin{tabular}{@{\hspace{2em}} ccc @{\hspace{2em}}} 
    Flavor & IC86 & IC93 \\
    \hline 
    $\nu_e+\bar{\nu}_e$        & 0.25\;mHz & 2.0\;mHz \\
    $\nu_\mu+\bar{\nu}_\mu$    & 0.76\;mHz & 4.3\;mHz \\
    $\nu_\tau+\bar{\nu}_\tau$  & 0.05\;mHz & 0.3\;mHz \\
    \end{tabular}
    \end{ruledtabular}
\end{table}

\begin{figure}[htb]
    \centering
    \vspace{-1em}
    \subfloat{
         \includegraphics[width=0.45\textwidth]{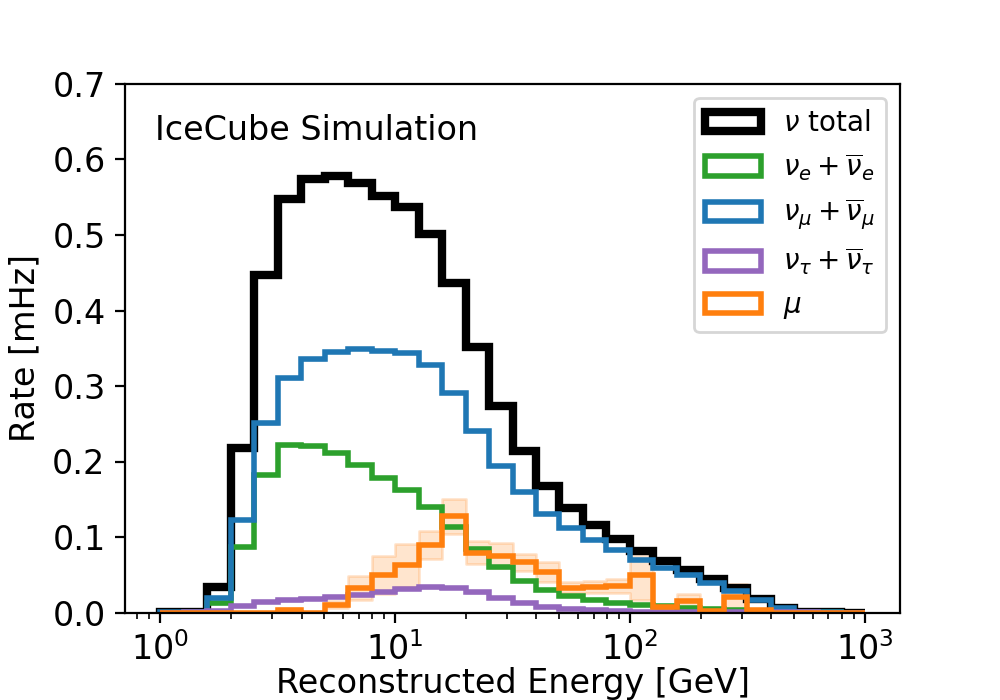}
         }
         
    \subfloat{
         \includegraphics[width=0.45\textwidth]{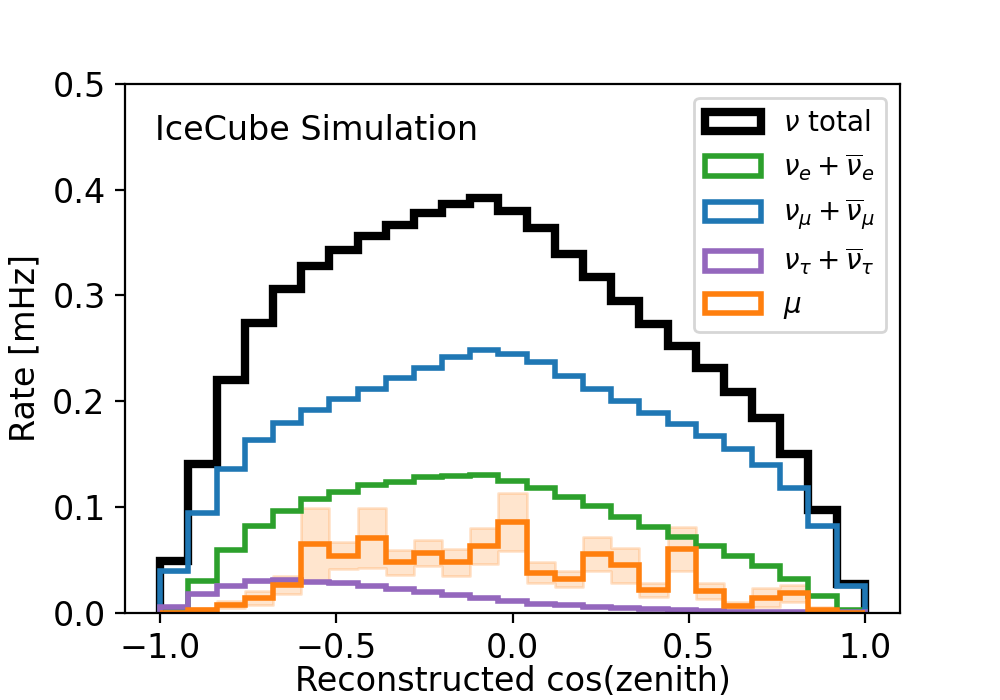}
         }
         
    \subfloat{
         \includegraphics[width=0.45\textwidth]{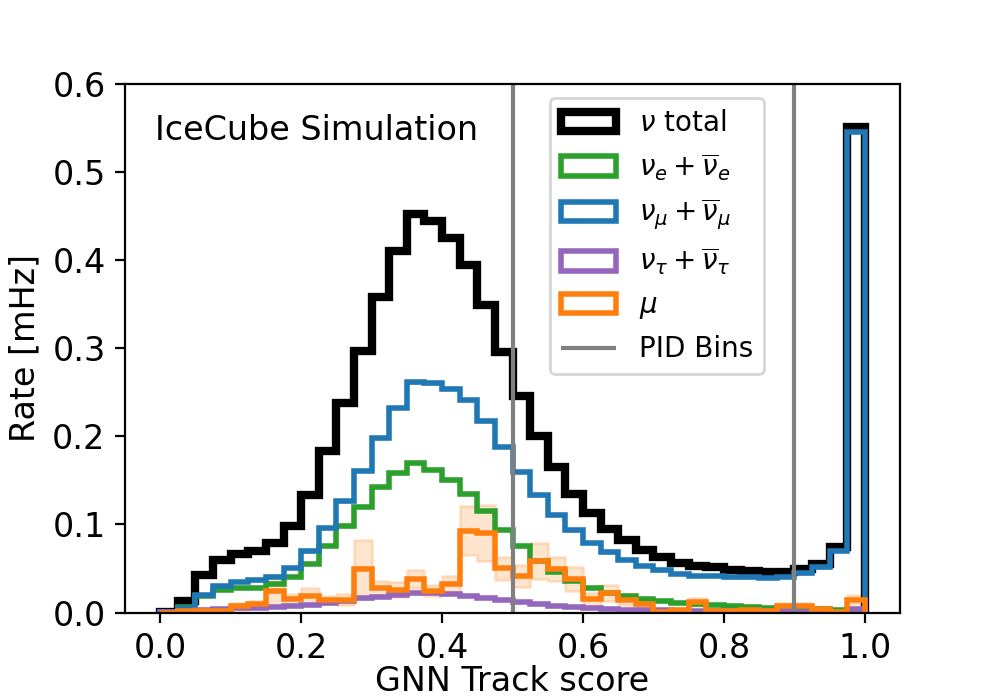}
         }
    \caption{Expected event distributions by flavor for the three variables used to bin events in the analysis: energy (top), cos($\theta_{zen}$) (middle), and particle identification score (bottom). In the bottom panel, a value of 1 indicates more track-like events and the grey lines correspond to the boundaries used for particle classification.}
    \label{fig:1d_events_by_flavor}
\end{figure}

Figure \ref{fig:1d_events_by_flavor} shows the expected event distributions by flavor after the above-mentioned cuts are applied for the three variables used to bin events at the analysis level: reconstructed energy, reconstructed cos(zenith), and the GNN track score indicating whether an event is more cascade-like or track-like. Note that the $\nu_e+\bar{\nu}_e$ component in the top panel peaks at reconstructed energies of about 3-5 GeV. These events are crucial in determining the neutrino mass ordering, and the lower the energy threshold of the IceCube Upgrade is key in achieving the increase in event rate in this region. This a major contributing factor to the improved NMO sensitivity described in more detail in Section \ref{sec:nmo}.

Finally, a few ``analysis level'' cuts on reconstructed energy and zenith are made to focus on the parameter space where most of the oscillation signal comes from and reduce some of the remaining muon background. This step requires a reconstructed energy between 3 and 300 GeV, and reconstructed cos(zenith) less than 0. At this level, the event sample is dominated by neutrinos, as can be seen in Fig.~\ref{fig:rates}. None of our simulated pure noise events survive after the final level cuts. Therefore, we show the 90\% upper limit of the pure noise rate based on the simulated livetime.

\subsection{Final Level Sample}
\label{sec:final_sample}

\begin{figure}[htb]
    \centering
    \includegraphics[width=\columnwidth]{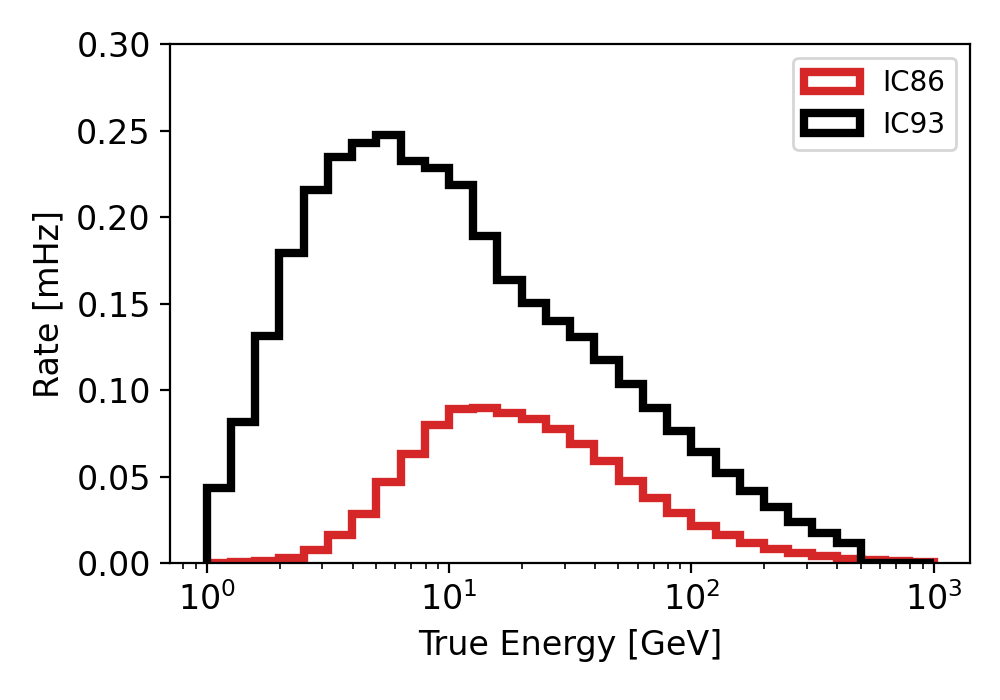}
    \caption{Comparison of the true energy distribution for simulated neutrino events in the IC86 and IC93 samples. The additional strings significantly increase the number of neutrinos detected.}
    \label{fig:dc_vs_upgrade_energy}
\end{figure}

Figure \ref{fig:dc_vs_upgrade_energy} shows the final level energy distributions after all cuts are made, and compares to the energy distribution for the IC86 sample. The energy distribution for IC93 peaks below 10 GeV, whereas for IC86 it peaks above 10 GeV. Additionally, the number of expected neutrino events increases across the entire energy range when using this selection.

\section{Analysis}
\label{sec:dc_icu}

\subsection{Analysis Set-Up}
\label{sec:combined_fit}

\begin{figure*}[htb]
    \centering
    \includegraphics[width=0.95\textwidth]{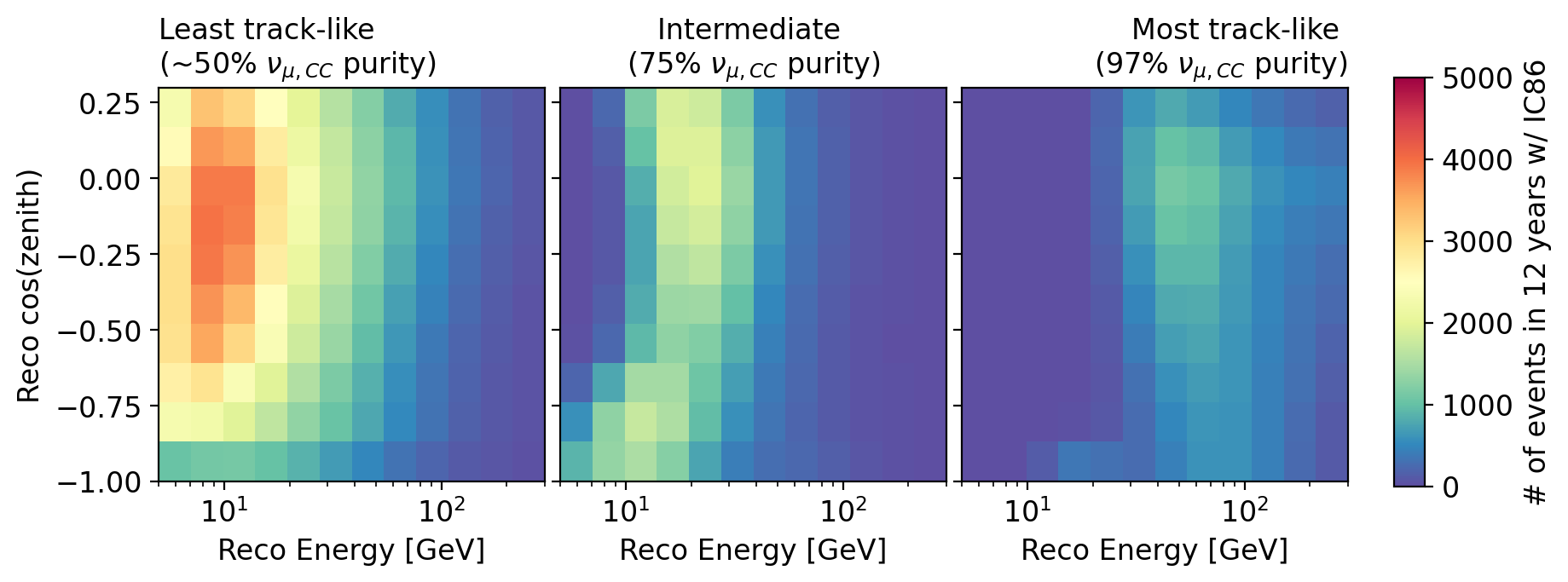}
    \vspace*{10pt}
    \includegraphics[width=0.95\textwidth]{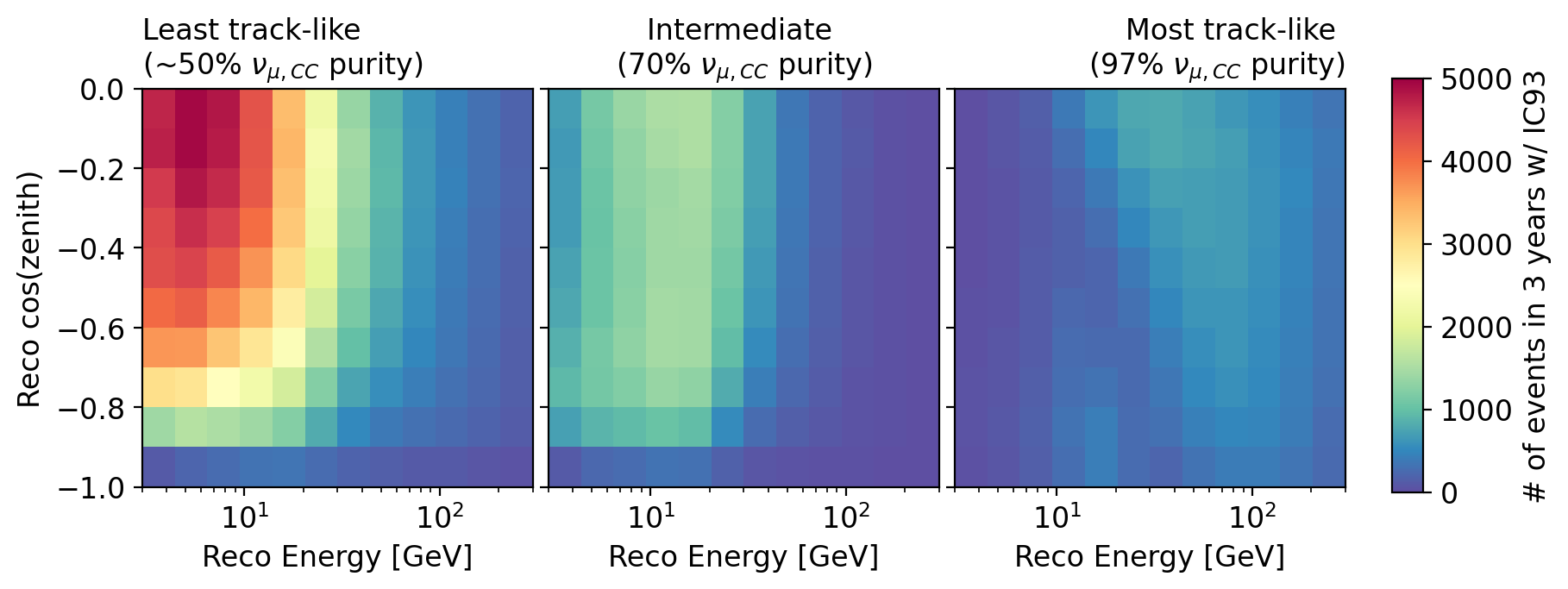}
    \caption{Distribution of events in the analysis binning used in this study. The top panel shows the events expected with 12 years of the current configuration (IC86), and the bottom panel shows the events expected with 3 years with the additional strings (IC93). Events are binned in energy, cos(zenith), and particle classification. The binning used in the top panel is the same configuration as existing IceCube DeepCore analyses, while the binning used for IC93 has been re-optimized for the new sample. The number of low energy events is increased with IC93 compared to IC86, even though it assumes fewer years of exposure. The sensitivities presented in this paper combine both of these datasets.}
    \label{fig:analysis_binning}
\end{figure*}

Once the additional strings are deployed there will already be approximately 12 years of data livetime with the current IC86 detector configuration. Therefore we use a combined analysis strategy to incorporate the existing years of IC86 data together with the anticipated IC93 data. Figure~\ref{fig:analysis_binning} shows the expected event distribution with 12 years of IC86 and with 3 years of IC93. The events are binned in energy, cos(zenith), and particle classification. 

The two simulated datasets share almost all simulation tools (Section~\ref{sec:particle}) and systematic parameters (Section~\ref{sec:systematics}) except for the efficiency of the new modules (DOM eff. ICU (IceCube Upgrade), but differ in the noise cleaning, reconstruction methods, event selection, and analysis binning. For IC86 we use the low-level processing described in~\cite{oscnext_prd}, the likelihood-based event reconstruction presented in~\cite{IceCube:2022kff}, and the analysis settings from~\cite{kayla_thesis}. The new tools discussed in (Sections~\ref{sec:detectorresponse}--\ref{sec:event_selection}) are applied only to IC93. The treatment of IC86 vs. IC93 datasets will be unified in the future.

For the sensitivity projections we compare two scenarios: one with no new hardware installed and data-taking continuing as it is today (denoted as ``IC86''). The other scenario includes the combined fit between 12 years of IC86 with additional years using the extra seven strings of the IceCube Upgrade from 2026 onwards (denoted as ``IC86 (12 yr) + IC93'').

To cover different possible scenarios, we show sensitivities at two different injection points. NuFit 5.2 with and without Super-Kamiokande (SK) data. 
We perform an Asimov analysis~\cite{Cowan:2010js} meaning that no pseudo-trials are generated to calculate sensitivities, but rather the expected neutrino counts in each analysis bin are used. This means that all sensitivities presented in this paper represent the median sensitivity.
A modified $\chi^{2}$, which takes into account uncertainties from MC statistics, is used as the test statistic:

\[\chi^2_{\mathrm{mod}}=\sum_{i}^{bins}\frac{(n_i - \mu_i)^2}{\mu_i + \sigma_i^2}+\sum_j^{syst}\frac{(\hat{s}_j-s_j)^2}{\sigma_j^2}\].

Here $n_i$ is the measured number of events in bin $i$, $\mu_i$ is the expected number of events, and $\sigma_i$ is the MC uncertainty. The second term represents the prior penalties for the systematic parameters $j$ which have Gaussian priors $\sigma_j$.

\subsection{Systematics Treatment}
\label{sec:systematics}

Atmospheric neutrino oscillation experiments have several major categories of systematic uncertainties that are incorporated as nuisance parameters in analyses. They arise from uncertainties in the atmospheric neutrino flux, the neutrino interaction cross-sections, and the detector itself. All systematics that were found to be important for the analyses presented in this paper are summarized in Table~\ref{table:sys}. These systematic uncertainties are informed by previous IceCube/DeepCore oscillation analyses.

\textbf{Flux:} The baseline atmospheric neutrino flux spectrum is the model by Honda \textit{et al.} \cite{Honda:2015fha}. Atmospheric neutrinos are created in hadronic decays after cosmic-ray interactions in the atmosphere. The uncertainty on the primary cosmic-ray spectrum is modeled with a power-law correction $E^{\Delta \gamma_\nu}$ \cite{Barr_2006}. 
To incorporate uncertainties on the hadronic interactions, the \textsc{MCEq} package \cite{Fedynitch:2018cbl}, which computes fluxes of atmospheric leptons, was used to calculate correction gradients that resemble the sources of systematic uncertainty from~\cite{Barr_2006}. Depending on the primary energy and the fraction of energy taken by the secondary meson, different uncertainties are used. While there are 17 parameters for the variations of the yields of $\pi^{\pm}$ and $K^{\pm}$, only 7 were found to be important for the analyses presented in this paper. They are listed in Table~\ref{table:sys}.

\textbf{Cross-section:} We simulate neutrinos with energies ranging from 1 GeV up to 500 GeV. Above $\sim$20 GeV, Deep Inelastic Scattering (DIS) is the dominant interaction type. We use \textsc{genie} \cite{genie_v2} and Cooper-Sarkar-Mertsch-Sarkar (CSMS) \cite{CSMS} to calculate the DIS cross-section and interpolate between the results. We use the parametrization described in \cite{oscnext_prd}, where DIS CSMS = 0 means that the \textsc{genie} value is used, and DIS CSMS = 1 would approximate the CSMS prediction. At lower energies, resonance production (RES), quasi-elastic scattering (QE), and coherent scattering (coh) become more important. In \textsc{genie} it is possible to re-weight the cross-sections of these interactions based on a change in the axial mass ($M_A$).

For the neutrino mass ordering analysis (see Section~\ref{sec:nmo}), we also include a parameter that scales between different $\nu_\tau$ cross-section models. A value of 0 indicates that the \textsc{genie} value is used; a non-zero value indicates a correction has been applied to re-weight events using a band that covers a range of models from \cite{Conrad:2010mh}. This parameter is not included in the $\nu_\tau$-norm study (see Section~\ref{sec:nutau_norm}) because the $\nu_\tau$ normalization can itself be interpreted as cross-section measurement; instead, the parameter is fixed to \textsc{genie}. \\

We have also studied the effect of differential DIS cross-section corrections to \textsc{genie} to better describe NuTeV data \cite{PhysRevD.74.012008}. These corrections are applied to the event weights, modeled as $a x^{-b}$ where $x$ is the Bjorken-x, and $a$ and $b$ are fit to the data. This was done independently for neutrinos and anti-neutrinos. The effective parameters were extracted using the NuTeV dataset at $E_\nu=35$\;GeV. Since a fit to the $E_\nu=65$\;GeV NuTeV data set yielded compatible parameter values, we assume no energy dependence. The corrections were found to have only a minor impact on the physics parameters and were mostly absorbed by our existing \textsc{genie} and flux systematics, and were therefore not explicitly included in the study.

In addition, we tested a similar procedure to model the impact of changes to the DIS Bodek-Yang correction~\cite{Bodek:2002vp}. Here the functional form $a x^{-b} y^{-c}$ was used where $x$ is the Bjorken-x and $y$ is the Bjorken-y or inelasticity. This parameterization showed similarly small impact on the physics parameters.

\textbf{Detector:} Uncertainties in the modeling of the optical properties of the ice and the efficiency of the optical modules have the greatest impact on our analyses. Our baseline ice model is described in \cite{paper_bfr}. Two scaling factors are used to vary the strength of the photon absorption and scattering in the ice, respectively. The optical properties of the refrozen ice in the immediate vicinity of the strings, which are altered by air bubbles and other impurities, are modeled following \cite{holeice_icrc2023}. This leads to two additional parameters ($p_0$, $p_1$) describing the modified angular efficiency of the optical modules. Note that this is only modeled for existing DOMs. During the deployment of the Upgrade, a modified drilling process is used to reduce air bubbles and other impurities. This is expected to improve the clarity of the refrozen ice surrounding the new modules. The uncertainty on the overall optical efficiency of the existing models (DOM eff. IC86) as well as the new modules (DOM eff. ICU) is taken into account. All new modules share the same efficiency scale, which has a tighter constraint than the one for the existing DOMs. Thanks to the new calibration devices, we assume the new modules to be calibrated with higher precision; the new modules have an uncertainty of 5\% and the existing modules have an uncertainty of 10\% as indicated in Table \ref{table:sys}.
To implement the effect of the detector systematics, for IC86 we use the ``hypersurface'' method described in \cite{oscnext_prd}, while for IC93 we use the likelihood-free inference method from \cite{Fischer:2023dbo}. In future studies, both detector configurations will use the same method.

\textbf{Normalization:} An overall normalization is applied to the neutrinos ($A_\mathrm{eff}$ scale) and the atmospheric muon background (Atm. $\mu$ scale) separately.

\textbf{Earth Model:} For the matter density profile of the Earth, the Preliminary Reference Earth Model \cite{prem} is used. For simplification, we assume 12 radial layers of fixed density, which does not lead to any additional free nuisance parameters.

\begin{table}[b!]
  \centering
  \caption{The parameters included in each of the analyses as physics and/or systematic nuisance parameters, along with the priors assumed for each. Parameters with Gaussian priors list the $\pm1\sigma$ range and parameters with uniform priors list the full allowed range. The last two columns indicate if the parameter is fit as a nuisance parameter in a given analysis (x for fixed, $\checkmark$ for free). A * after free indicates that parameter is the physics parameter of interest. The atmospheric oscillation parameters ($\theta_{23},\Delta m^2_{32}$) and tau neutrino normalization ($\nu_\tau$) analyses use the same list of free/fixed nuisance parameters, whereas the neutrino mass ordering (NMO) analysis has its own set.}
  \label{table:sys}
    \begin{ruledtabular}
    \begin{tabular}{lccccc} 
    Parameter  & Nominal & Prior & $\theta_{23},\Delta m^2_{32}$ & $\nu_\tau$ & NMO \\ 
    \hline
    \textbf{Detector:}        &  & &  \\
    DOM eff. IC86   & 1.0  & $\pm$0.1  & $\checkmark$ & $\checkmark$ & $\checkmark$ \\
    DOM eff. ICU    & 1.0  & $\pm$0.05 & $\checkmark$ & $\checkmark$ & $\checkmark$ \\
    Ice absorption  & 1.0  & $\pm$0.05 & $\checkmark$ & $\checkmark$ & $\checkmark$ \\
    Ice scattering  & 1.0  & $\pm$0.1  & $\checkmark$ & $\checkmark$ & $\checkmark$ \\ 
    Relative eff. $p_0$  & 0.10  &  [-0.6, 0.5] & $\checkmark$ & $\checkmark$ & $\checkmark$ \\
    Relative eff. $p_1$  & -0.05  & [-0.15, 0.05] & $\checkmark$ & $\checkmark$ & $\checkmark$ \\
    \hline
    \textbf{Flux:}  &     &    &    & \\
    $\Delta \gamma_\nu$         & 0.0  & $\pm$0.1  & $\checkmark$ & $\checkmark$ & $\checkmark$ \\
    $\Delta \pi^\pm$ yields D   & 0.0  & $\pm$0.3  & x & x & $\checkmark$ \\
    $\Delta \pi^\pm$ yields G   & 0.0  & $\pm$0.3  & $\checkmark$ & $\checkmark$ & $\checkmark$ \\
    $\Delta \pi^\pm$ yields H   & 0.0  & $\pm$0.15 & $\checkmark$ & $\checkmark$ & $\checkmark$ \\
    $\Delta \pi^\pm$ yields I   & 0.0  & $\pm$0.61 & $\checkmark$ & $\checkmark$ & $\checkmark$ \\
    $\Delta K^+$ yields W       & 0.0  & $\pm$0.4 & $\checkmark$ & $\checkmark$ & $\checkmark$ \\
    $\Delta K^+$ yields Y       & 0.0  & $\pm$0.3 & $\checkmark$ & $\checkmark$ & x \\
    $\Delta K^+$ yields Z       & 0.0  & $\pm$0.122 & x & x & $\checkmark$ \\ 
    \hline
    \textbf{Cross-section:}   &     &    &    & \\
    $M_{A}^{CCQE}$ (in $\sigma$)    & 0.0 & $\pm$1.0 & $\checkmark$ & $\checkmark$ & $\checkmark$ \\
    $M_{A}^{CCRES}$ (in $\sigma$)   & 0.0 & $\pm$1.0 & $\checkmark$ & $\checkmark$ & $\checkmark$ \\
    $M_{A}^{NCRES}$ (in $\sigma$)   & 0.0 & $\pm$1.0 & $\checkmark$ & $\checkmark$ & $\checkmark$ \\
    $M_{A}^{coh}$ (in $\sigma$)     & 0.0 & $\pm$1.0 & x & x & $\checkmark$ \\
    DIS CSMS                        & 0.0 & $\pm$1.0 & $\checkmark$ & $\checkmark$ & x \\
    $\nu_\tau$ x-sec scale          & 0.0 & [-1.0, +1.0] & x & x & $\checkmark$ \\
    \hline
    \textbf{Normalization:}   &     &    &    & \\
    $A_\mathrm{eff}$ scale  & 1.0 & [0.1, 2.0]  & $\checkmark$ & $\checkmark$ & $\checkmark$ \\
    \hline
    \textbf{Atm. muons:}        &  &  & &  \\
    Atm. $\mu$ scale & 1.0 & [0.1, 3.] & $\checkmark$ & $\checkmark$ & $\checkmark$ \\ 
    \hline
    \textbf{Oscillations:}   &     &    &    & \\
    $\theta_{13}$            & NuFit5.2  & 0.11 & x & x & $\checkmark$ \\
    $\theta_{23}$            & NuFit5.2  & None & $\checkmark$* & $\checkmark$ & $\checkmark$ \\
    $\Delta m^2_{32}$        & NuFit5.2  & None & $\checkmark$* & $\checkmark$ & $\checkmark$ \\
    $\nu_\tau$ normalization & 1.0  & None & x & $\checkmark$* & x \\
    mass ordering            & normal   & n/a  & x & x & $\checkmark$* \\
    \hline
    \end{tabular}
    \end{ruledtabular}
\end{table}

There are more systematics that are being targeted as part of the Upgrade ice calibration, such as the exact geometry/position of the modules after deployment, the optical properties of the refrozen ice directly surrounding the modules, uncertainties on the depth and direction-dependent photon scattering function. They are not considered in this paper, but will be studied after the deployment using the calibration devices mentioned in Section~\ref{sec:calibration_modules}.

\section{Expected Sensitivities}
\label{sec:sensitivities}

\subsection{\texorpdfstring{Sensitivity to $\theta_{23}$ and $\Delta m^2_{32}$}{}}
\label{sec:theta23_deltam2_32}

The mixing between neutrino states is very strong, with $\theta_{23}$ being close to maximal mixing. This makes it important to measure the neutrino oscillation parameters, especially $\theta_{23}$, because they can reveal structure in the weak flavor mixing of leptons.
To obtain the expected constraints on the atmospheric oscillation parameters $\Delta m^2_{32}$ and $\theta_{23}$, we follow a similar analysis procedure as for the existing IceCube measurements \cite{oscnext_prd,IceCube:2024oscnext}.

\begin{figure*}[ht]
     \centering
     \vspace{-1em}
     \subfloat{
         \includegraphics[width=0.45\textwidth]{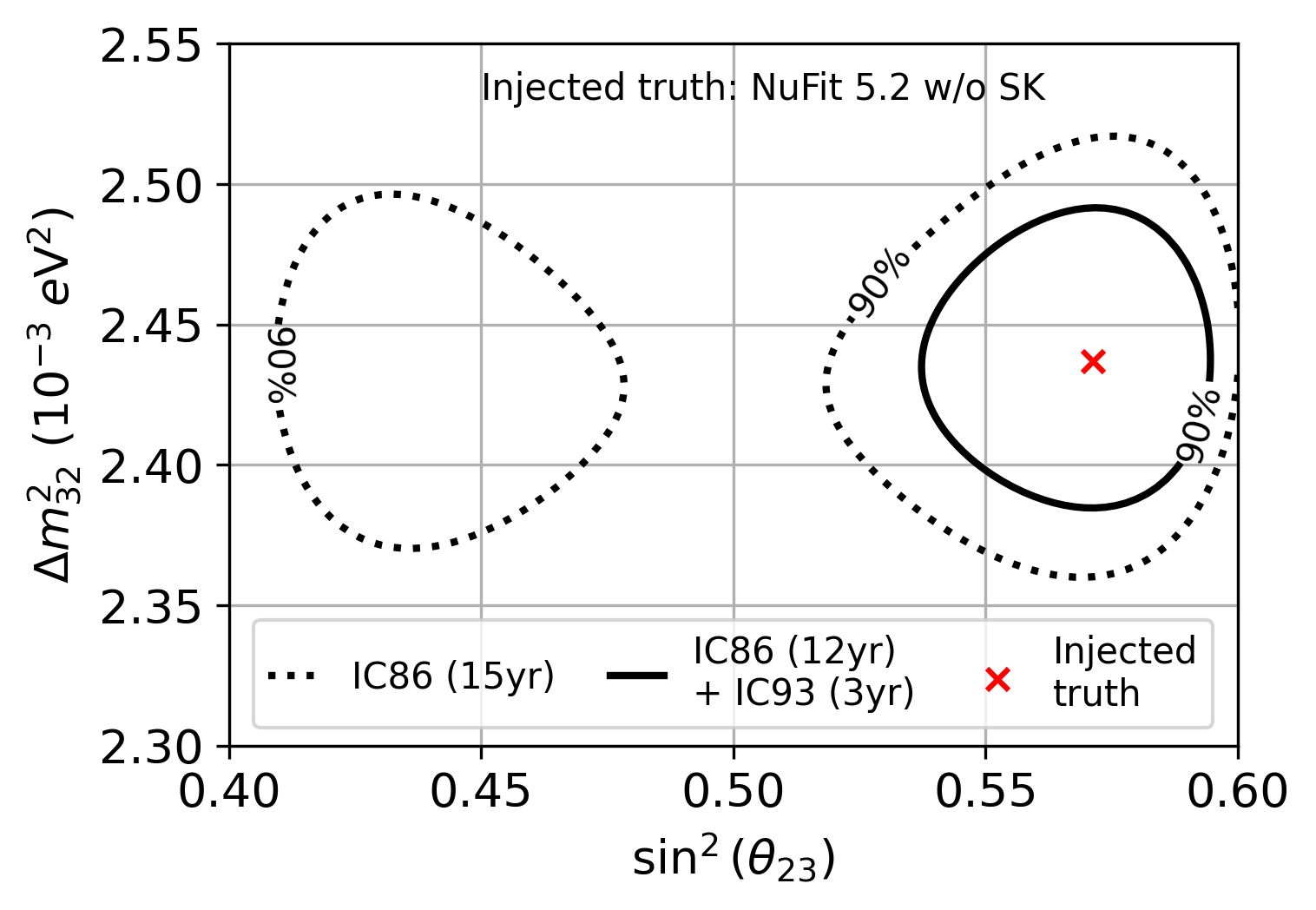}
         }
     \subfloat{
         \includegraphics[width=0.45\textwidth]{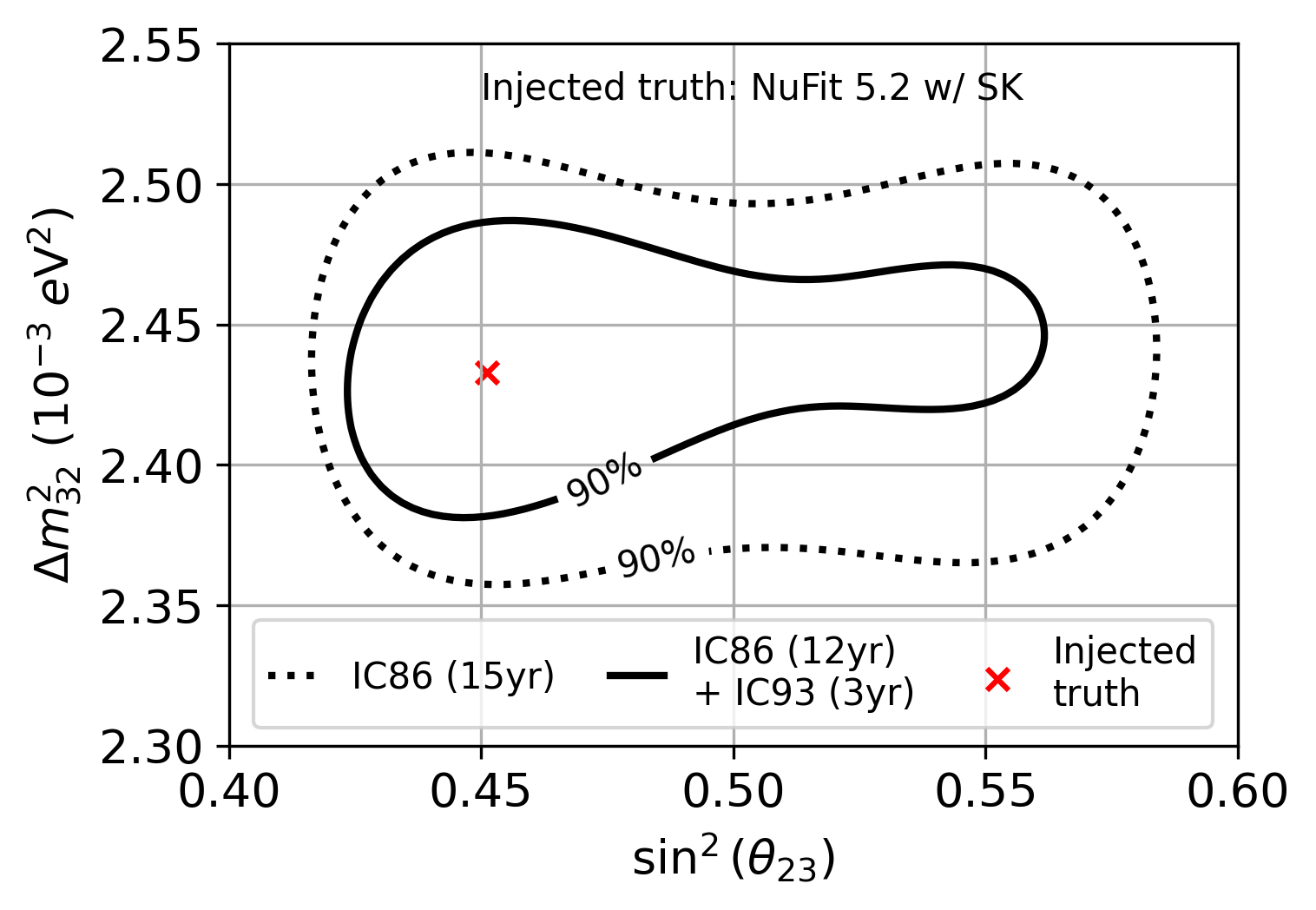}
         }
    \caption{Expected 90\% confidence level sensitivity contours for the atmospheric oscillation parameters 3 years after the planned deployment of the additional strings. A scenario w/ (solid) and w/o (dotted) the additional strings is compared in each plot. The left plot uses NuFit 5.2 w/o SK as injected truth, while the right plot assumes NuFit 5.2 w/ SK.} 
    \label{fig:numu_ic86_vs_ic93}
\end{figure*}

Figure~\ref{fig:numu_ic86_vs_ic93} shows the sensitivity at the 90\% confidence level with 3 years of IC93 plus 12 years of IC86 data. To illustrate the sensitivity gain from the additional strings, we also show a scenario with 15 years of data from the current detector configuration (IC86, dotted lines). The left plot assumes NuFit 5.2 \cite{nufit_5_2} (w/o SK data) as injected truth, while the right plot uses NuFit 5.2 (w/ SK).

Thanks to the additional strings, IceCube's 90\% C.L. region for $\Delta m^2_{32}$ and $\theta_{23}$ shrinks by about 70\% (w/o SK) and 55\% (w/ SK). IceCube's ability to rule out maximal mixing or determine the octant strongly depends on the true value of $\theta_{23}$. As seen in Fig.~\ref{fig:numu_ic86_vs_ic93}, for an injected truth given by the NuFit 5.2 w/o SK value, we could exclude maximal mixing at the 90\% confidence level, and the addition of the IceCube Upgrade strings could allow us to rule out the lower octant. However, for other injected truth values like NuFit 5.2 w/ SK value, the octant is not resolved and maximal mixing is not able to be rejected at the 90\% confidence level. However this study was not specifically optimized for resolving the octant or rejecting maximal mixing so future work could improve IceCube's sensitivity to these questions.  

\begin{figure*}[ht]
     \centering
     \vspace{-1em}
     \subfloat{
         \includegraphics[width=0.45\textwidth]{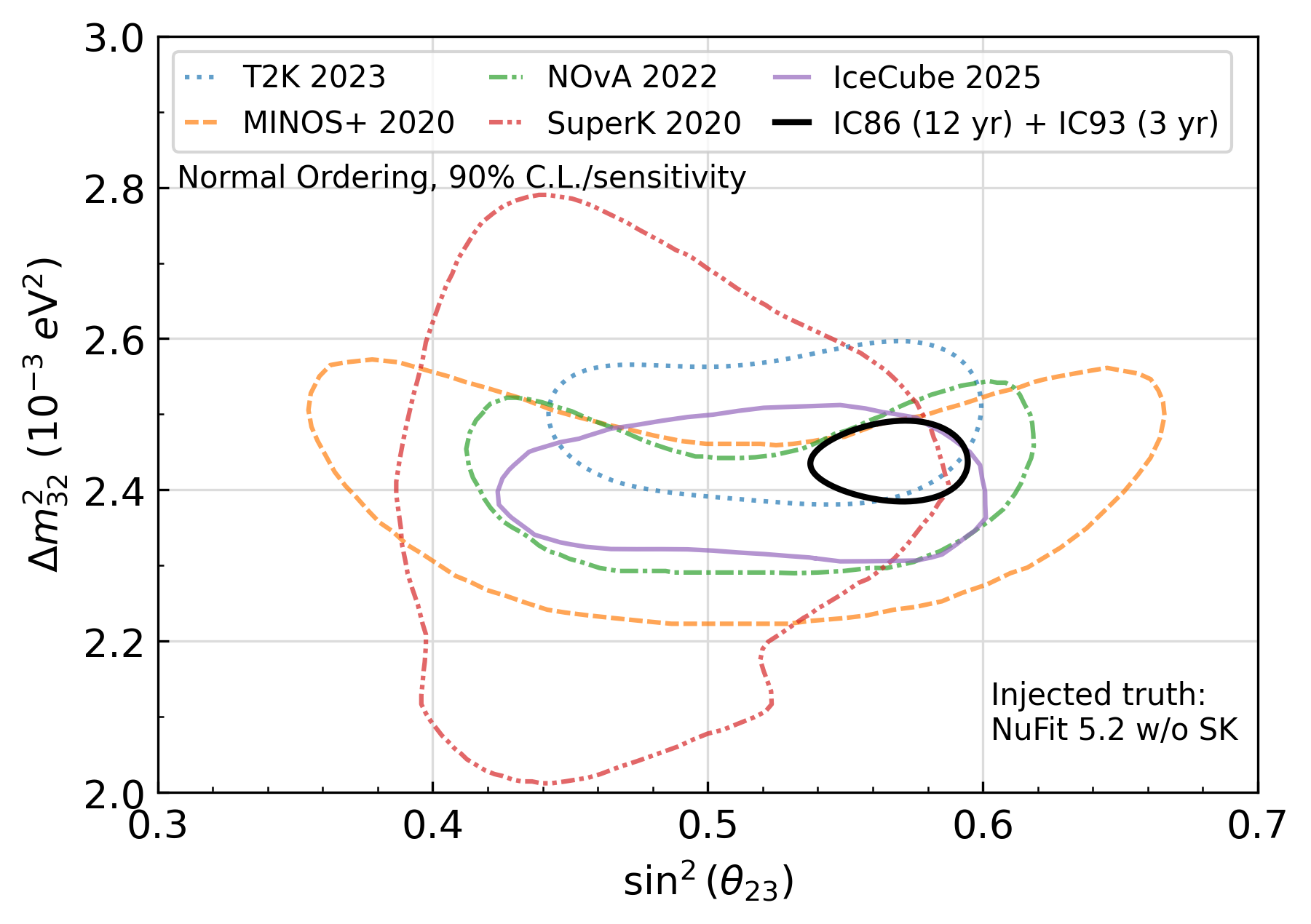}
         }
     \subfloat{
         \includegraphics[width=0.45\textwidth]{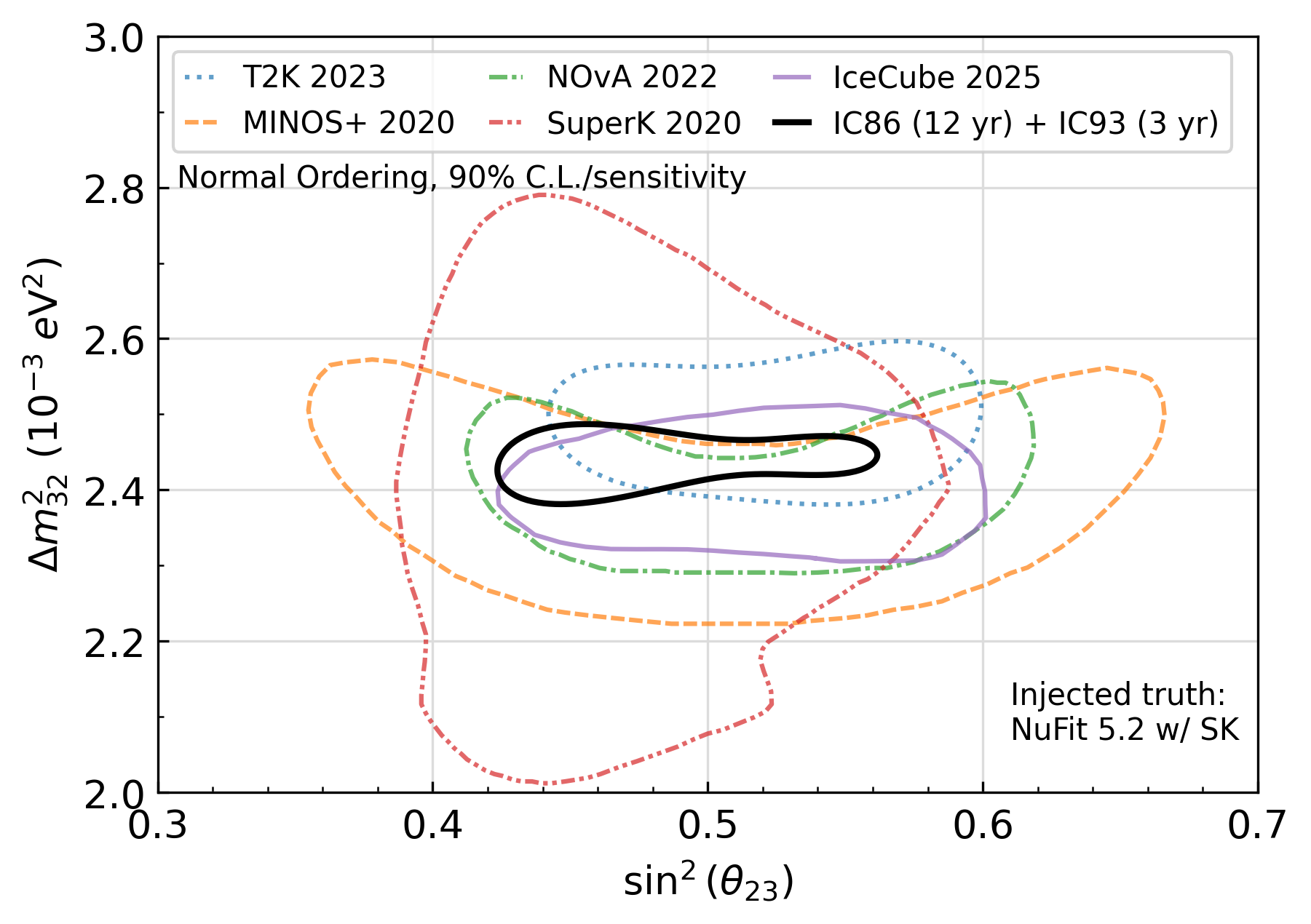}
         }
    \caption{The expected IceCube 90\% confidence level sensitivity contour with 3 years of livetime including the additional strings compared to existing measurements of the atmospheric oscillation parameters. The left plot uses NuFit 5.2 w/o SK as injected truth, while the right plot assumes NuFit 5.2 w/ SK. Results from several experiments are shown for comparison \cite{IceCube:2024oscnext,T2K:2023smv,NOvA:2021nfi,MINOS:2020llm,SuperK}.}
    \label{fig:numu_ic93_vs_other_experiments}
\end{figure*}

Appendix~\ref{sec:appendix_analysis} shows one-dimensional projections of these sensitivities, as well as the two-dimensional contour for an additional scenario where the injected truth is given by \cite{IceCube:2024oscnext}. 
Our projected 3 year sensitivities compared to current experimental measurements \cite{IceCube:2024oscnext,T2K:2023smv,NOvA:2021nfi,MINOS:2020llm,SuperK} are shown in Fig.~\ref{fig:numu_ic93_vs_other_experiments}.

\subsection{\texorpdfstring{Sensitivity to $\nu_{\tau}$ Normalization}{}}
\label{sec:nutau_norm}

\begin{figure*}[htb]
    \centering
    \vspace{-1em}
     \subfloat{
         \includegraphics[width=0.42\textwidth]{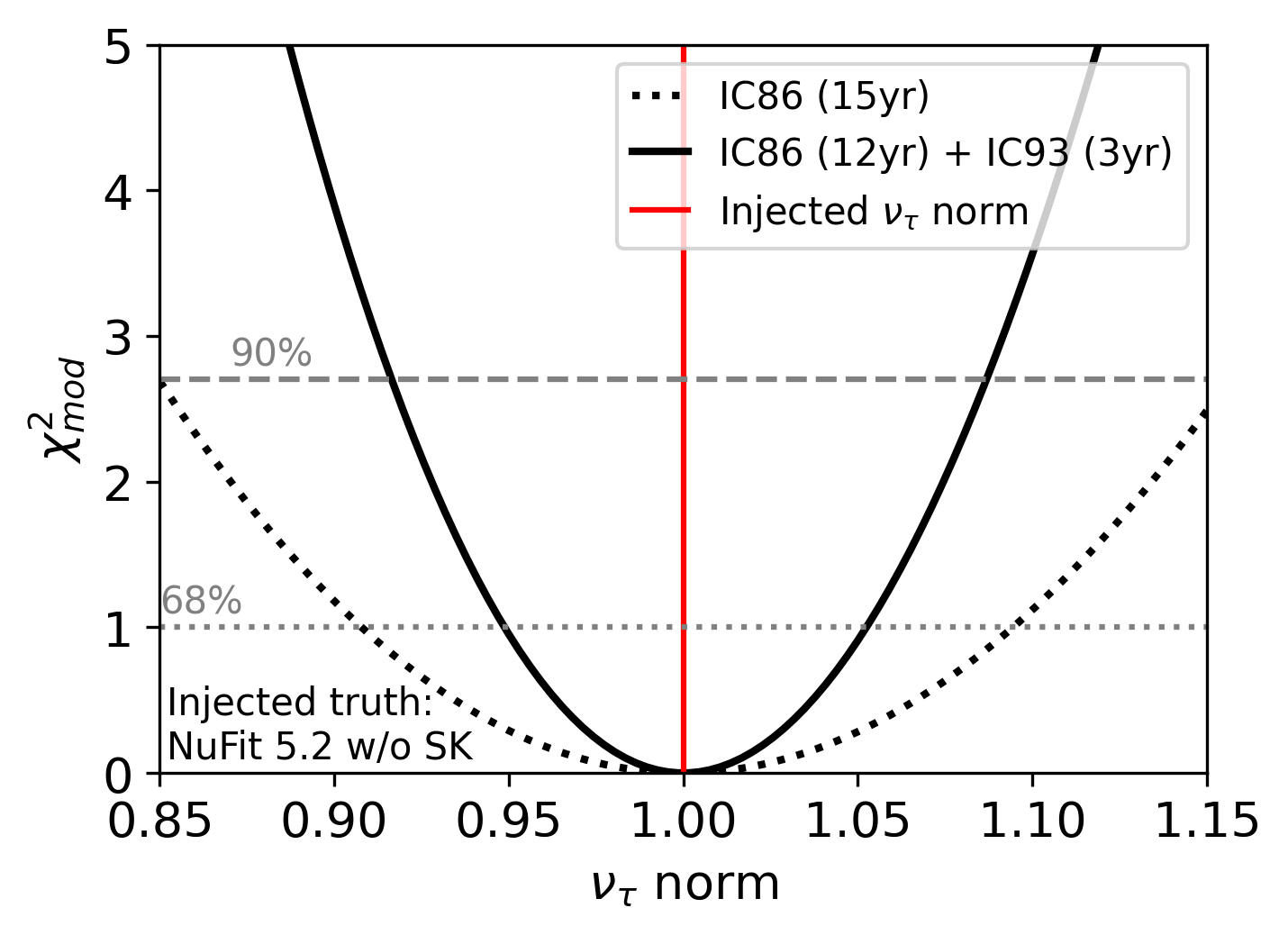}
         }
     \subfloat{
         \includegraphics[width=0.46\textwidth]{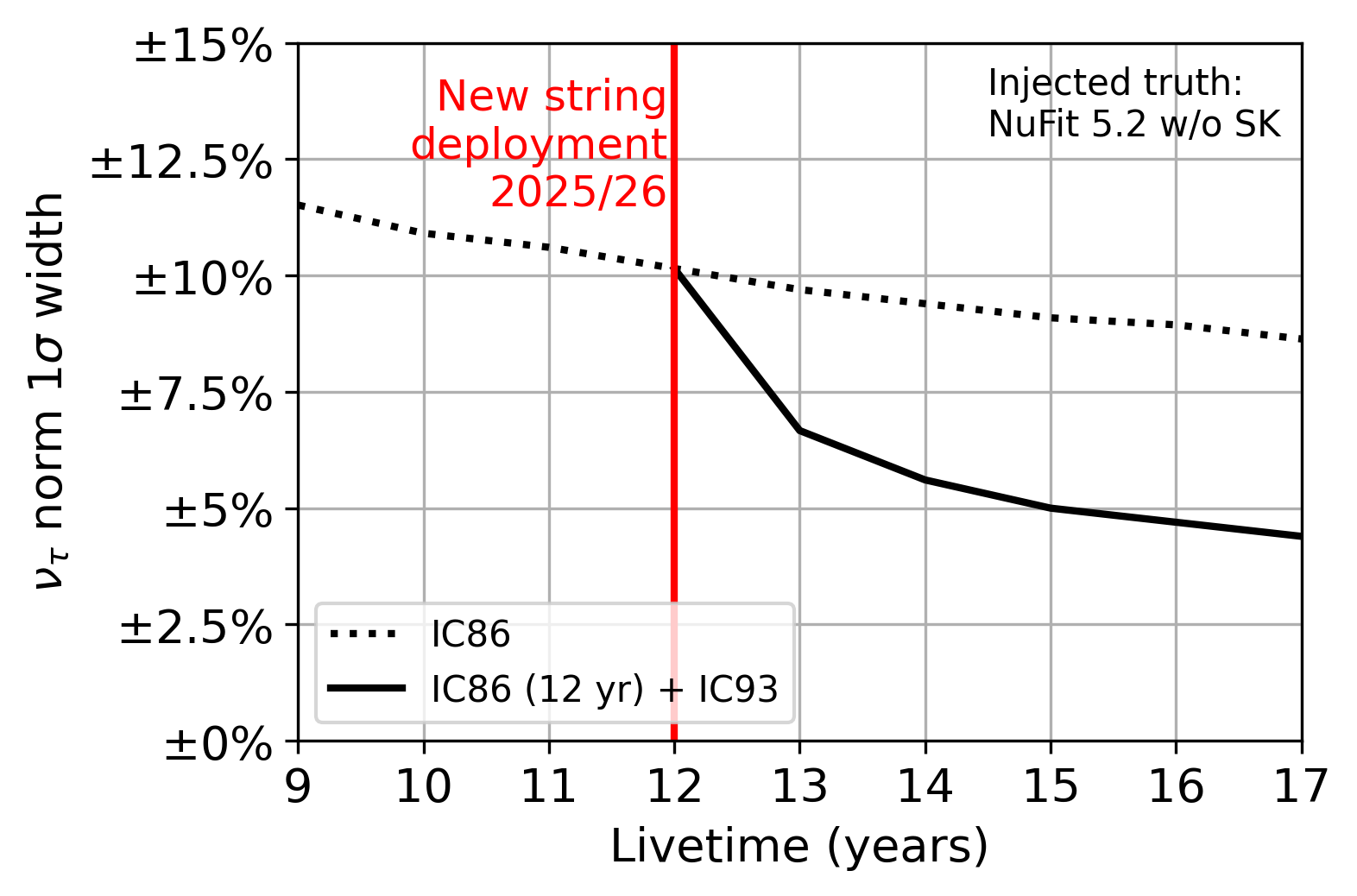}
         }
    \caption{Expected sensitivity for constraining the $\nu_{\tau}$ normalization scaling factor. The left plot shows a profile scan 3 years after the planned deployment of the additional strings, while the right plot shows the 1$\sigma$ width of such a scan over detector livetime. In both plots a scenario  w/ (solid) and w/o (dotted) the additional strings is compared.}
    \label{fig:nutau}
\end{figure*}

We will also measure the overall rate of $\nu_\tau$ and compare it to the oscillation expectation. This is done by scaling the number of observed $\nu_\tau$ by a normalization factor and can be interpreted as test of the unitarity of the PMNS matrix as well as a test of the $\nu_\tau$ cross-section. A normalization value of 1.0 indicates the nominal expectation. A value other than 1.0 could hint at new physics. More information about how IceCube measures $\nu_\tau$ appearance can be found in \cite{IceCube:2019nutauprd}.

Figure~\ref{fig:nutau} illustrates our sensitivity to the $\nu_{\tau}$ normalization for a true value of 1.0. The left plot shows the rejection power for different $\nu_{\tau}$ normalization values assuming data from 3 years after the planned deployment, while in the right plot the 1$\sigma$ uncertainty on $\nu_{\tau}$ normalization is shown for different detector operation times. Since the choice of injected oscillation parameters has a negligible impact on this analysis, we only show the results for NuFit 5.2 w/o SK.

Already after a few years the uncertainty on the $\nu_{\tau}$ normalization can be reduced by almost a factor of two compared to the configuration without the additional strings. A precision of 5\% at 1$\sigma$ is expected with 3 years of livetime. Future studies could constrain non-unitarity more directly by fitting the PMNS matrix elements, similar to the method in \cite{Kozynets:2024xgt}.

\subsection{Sensitivity to Neutrino Mass Ordering}
\label{sec:nmo}

This analysis covers our sensitivity to the ordering of the three neutrino mass eigenstates, which can either be ``normal" or ``inverted". Recent cosmological measurements of baryonic acoustic oscillations provide an upper limit on the sum of neutrino masses that is remarkably close to the minimum allowed value from neutrino oscillations for the inverted ordering. Therefore, a direct measurement of the NMO is of great importance: the inverted ordering could be an indication of the need to adjust cosmological models or the reconsideration of standard neutrino oscillation assumptions~\cite{Jiang:2024viw}.

All sensitivities refer to median sensitivities as defined in \cite{Blennow:2013oma}. More information about how IceCube determines the NMO can be found in \cite{IceCube:2019dyb}.
As shown in Table~\ref{table:sys}, the selection of free parameters was adjusted for this analysis, because the signal region for this analysis is at lower energies compared to the analyzes presented in Sections~\ref{sec:theta23_deltam2_32} and \ref{sec:nutau_norm}. 

\begin{figure*}[htb]
     \centering
     \vspace{-1em}
     \subfloat{
         \includegraphics[width=0.44\textwidth]{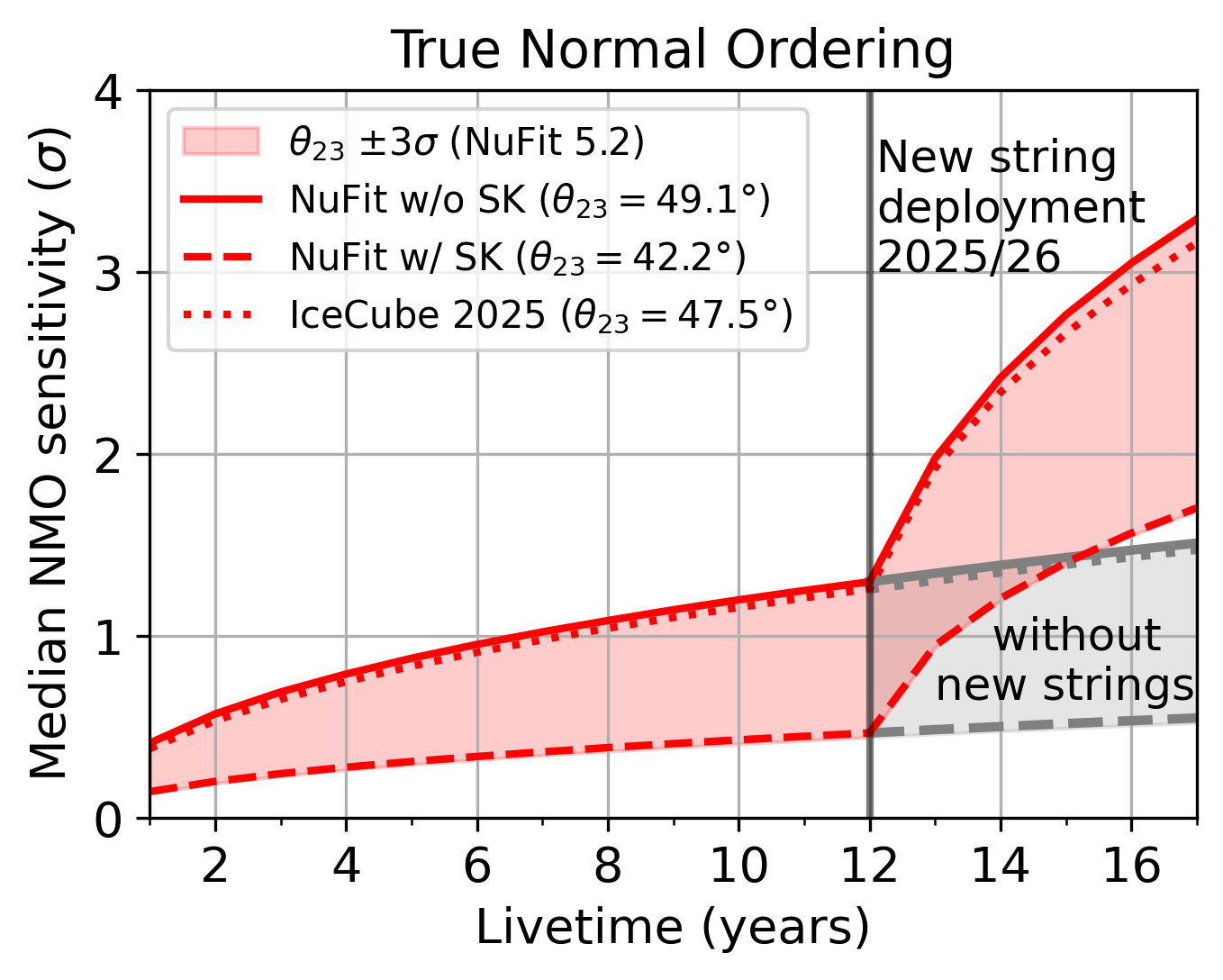}
         }
     \subfloat{
         \includegraphics[width=0.46\textwidth]{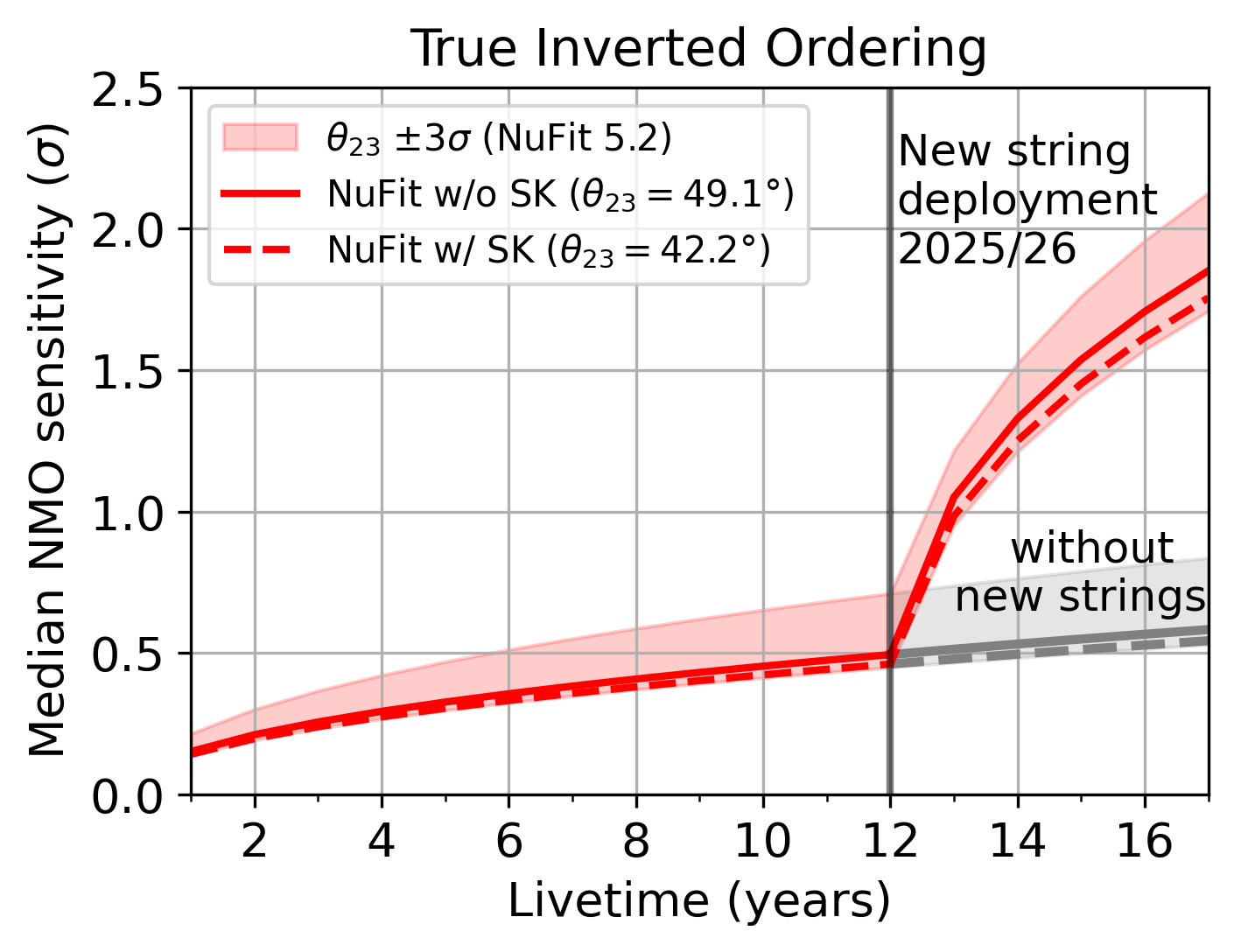}
         }
    \caption{Livetime evolution of the median NMO sensitivity for different true values of $\theta_{23}$. The shaded region marks the possible range of sensitivities for the NuFit 5.2 3$\sigma$ range of $\theta_{23}$. The left plot is for a true normal ordering, and the right plot for a true inverted ordering.}
    \label{fig:nmo}
\end{figure*}

Figure~\ref{fig:nmo} shows the expected median NMO sensitivity evolution as a function of the detector livetime assuming a true normal ordering (left plot) or a true inverted ordering (right plot). Since the NMO sensitivity of atmospheric neutrino experiments strongly depends on the true value of $\theta_{23}$, Fig.~\ref{fig:nmo} does not provide a single median sensitivity but a range (shaded areas) of possible sensitivities depending on the assumed true value of $\theta_{23}$. In addition, we show sensitivities for three example values of $\theta_{23}$: the value from NuFit 5.2 w/o SK (solid line), NuFit 5.2 w/ SK (dashed line), and the IceCube result from \cite{IceCube:2024oscnext} (dotted line). In contrast to accelerator long-baseline experiments, a CP violating phase $\delta_{CP}$ in the mixing matrix has negligible impact on our NMO sensitivities. This is mainly due to the higher energy range of atmospheric neutrinos.

The sensitivity improvement from the additional strings can be seen by comparing the grey and red sensitivities right of the vertical grey line, which marks the deployment of the additional strings. For a true normal ordering a detection significance of up to $3\sigma$ is possible within 5 years, while for a true inverted ordering up to $2\sigma$ is possible.

\section{Conclusion}
\label{sec:conclusion}

Seven additional strings, hosting newly developed optical modules, will be deployed for the IceCube Upgrade in the Polar season of 2025-2026. These strings will significantly increase the number of GeV neutrino interactions observed by the detector, as well as the ability to reconstruct and classify events. These new strings will therefore lead to significant improvement in our sensitivities to many neutrino oscillation analyses, particularly the three analyses considered in this paper. With three years of data from the upgraded detector, we expect to collect about $3.3\times10^5$ neutrino events with the event selection presented here. Combining the new data with the approximately $2.8\times10^5$ neutrinos expected from the 12 year data taken by the current IceCube DeepCore, yields the sensitivities presented in this paper. Compared to a scenario without the additional strings, we see an improvement of 55-70\% in the area enclosed by the 90\% C.L. contour for the atmospheric neutrino oscillation parameters, around 40\% in the 1$\sigma$ range for the $\nu_\tau$ normalization, and a factor of $2-3\times$ boost in the median NMO sensitivity.

Additional optimizations of triggers, event selection, processing, reconstructions, analysis choices, and models for treating systematic uncertainties are expected to further improve the sensitivities presented here. In addition, opportunities for future improvements in sensitivity can be achieved through combined fits with other experiments. For example, previous studies have shown that a combined fit between the IceCube Upgrade and a medium baseline reactor neutrino experiment like Jiangmen Underground Neutrino Observatory (JUNO) \cite{JUNO:2015zny} will significantly increase the joint NMO sensitivity \cite{juno_icecube}.

Beyond atmospheric oscillations, many additional analyses will also benefit from the reduced energy threshold, higher event rate, and improved precision of the IceCube Upgrade. Examples for such analyses are dark matter searches, beyond the standard model searches, and GeV neutrino astronomy. Finally, the extensive suite of calibration devices that will be deployed in the Upgrade will lead to better knowledge of the ice properties and reduce the impact of detector systematic uncertainty across all IceCube analyses. The IceCube Upgrade is expected to bring unprecedented precision to the atmospheric neutrino sector.

\begin{acknowledgements}
The authors gratefully acknowledge the support from the following agencies and institutions:
USA {\textendash} U.S. National Science Foundation-Office of Polar Programs,
U.S. National Science Foundation-Physics Division,
U.S. National Science Foundation-EPSCoR,
U.S. National Science Foundation-Office of Advanced Cyberinfrastructure,
Wisconsin Alumni Research Foundation,
Center for High Throughput Computing (CHTC) at the University of Wisconsin{\textendash}Madison,
Open Science Grid (OSG),
Partnership to Advance Throughput Computing (PATh),
Advanced Cyberinfrastructure Coordination Ecosystem: Services {\&} Support (ACCESS),
Frontera and Ranch computing project at the Texas Advanced Computing Center,
U.S. Department of Energy-National Energy Research Scientific Computing Center,
Particle astrophysics research computing center at the University of Maryland,
Institute for Cyber-Enabled Research at Michigan State University,
Astroparticle physics computational facility at Marquette University,
NVIDIA Corporation,
and Google Cloud Platform;
Belgium {\textendash} Funds for Scientific Research (FRS-FNRS and FWO),
FWO Odysseus and Big Science programmes,
and Belgian Federal Science Policy Office (Belspo);
Germany {\textendash} Bundesministerium f{\"u}r Bildung und Forschung (BMBF),
Deutsche Forschungsgemeinschaft (DFG),
Helmholtz Alliance for Astroparticle Physics (HAP),
Initiative and Networking Fund of the Helmholtz Association,
Deutsches Elektronen Synchrotron (DESY),
and High Performance Computing cluster of the RWTH Aachen;
Sweden {\textendash} Swedish Research Council,
Swedish Polar Research Secretariat,
Swedish National Infrastructure for Computing (SNIC),
and Knut and Alice Wallenberg Foundation;
European Union {\textendash} EGI Advanced Computing for research;
Australia {\textendash} Australian Research Council;
Canada {\textendash} Natural Sciences and Engineering Research Council of Canada,
Calcul Qu{\'e}bec, Compute Ontario, Canada Foundation for Innovation, WestGrid, and Digital Research Alliance of Canada;
Denmark {\textendash} Villum Fonden, Carlsberg Foundation, and European Commission;
New Zealand {\textendash} Marsden Fund;
Japan {\textendash} Japan Society for Promotion of Science (JSPS)
and Institute for Global Prominent Research (IGPR) of Chiba University;
Korea {\textendash} National Research Foundation of Korea (NRF);
Switzerland {\textendash} Swiss National Science Foundation (SNSF).
\end{acknowledgements}

\bibliography{references}

\appendix
\clearpage
\section{Additional Sensitivities}
\label{sec:appendix_analysis}

\begin{figure}[ht]
     \subfloat{
         \includegraphics[width=0.46\textwidth]{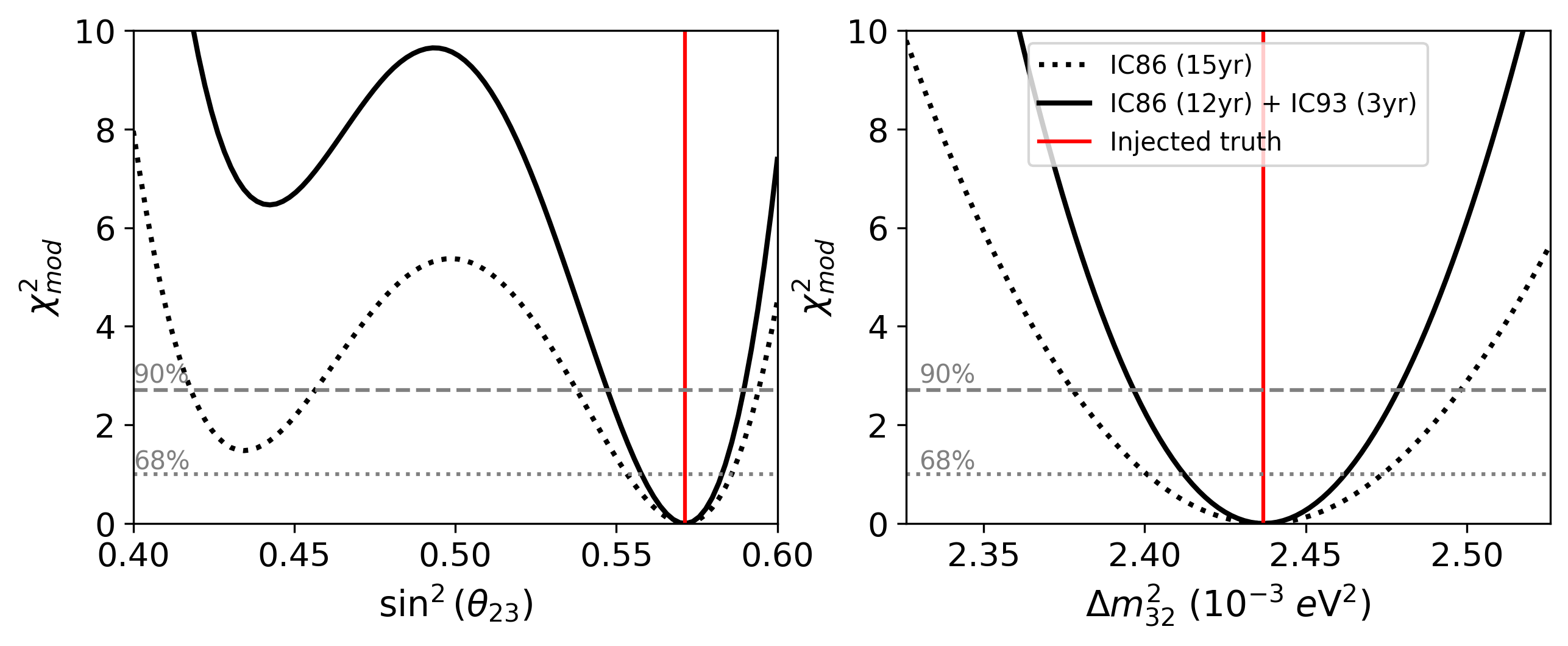}
         }
     
     \subfloat{
         \includegraphics[width=0.46\textwidth]{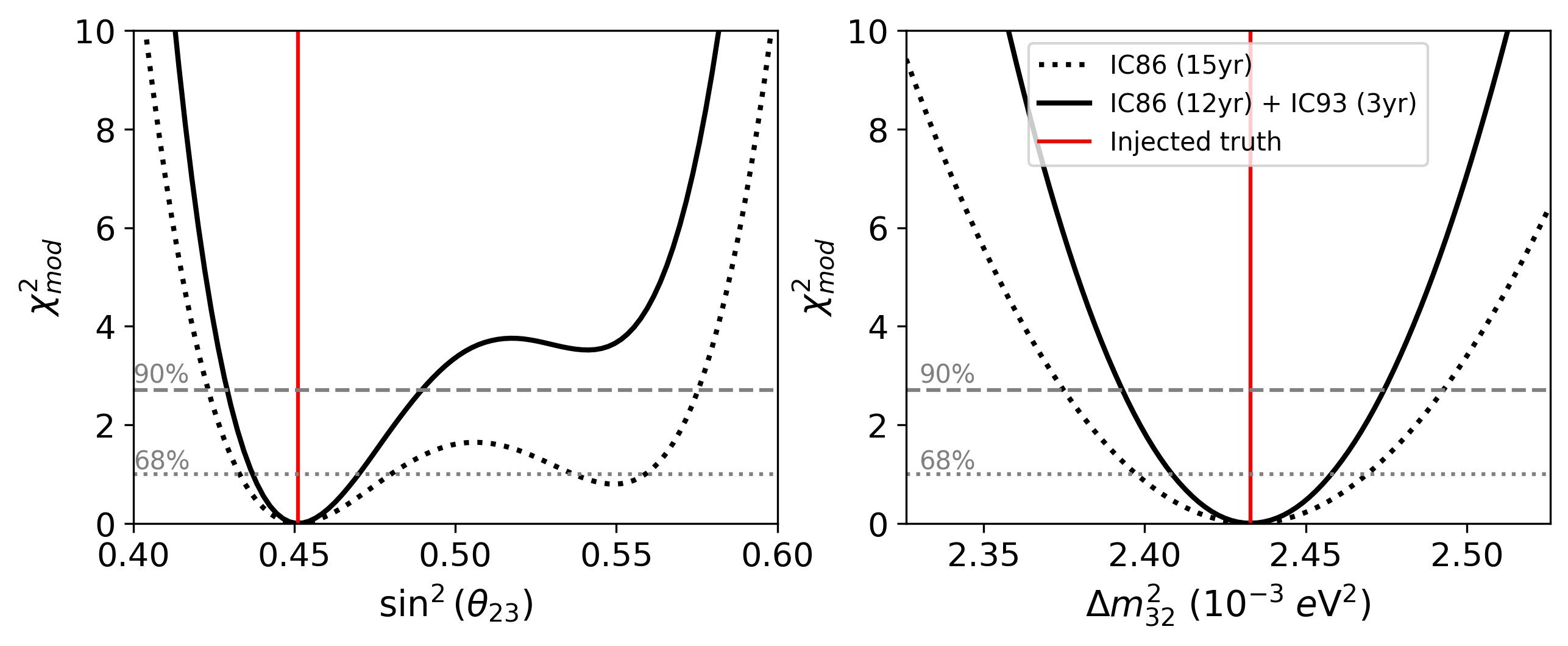}
         }
    \caption{Expected sensitivity to the atmospheric oscillation parameters with 3 years of livetime with the new strings. A scenario w/ (solid) and w/o (dotted) the additional strings is compared in each plot. The top plots use NuFit 5.2 w/o SK as injected truth, while the bottom plots assume NuFit 5.2 w/ SK.}
    \label{fig:numu_1d}
\end{figure}

Figure \ref{fig:numu_1d} shows the 1D projections of the expected sensitivity to the atmospheric oscillation parameters sin$^2(\theta_{23})$ and $\Delta m^2_{32}$. These correspond to the 2D projections shown in Figs. \ref{fig:numu_ic86_vs_ic93} and \ref{fig:numu_ic93_vs_other_experiments}. 

\begin{figure}[b!]
     \centering
     \subfloat{
         \includegraphics[width=0.45\textwidth]{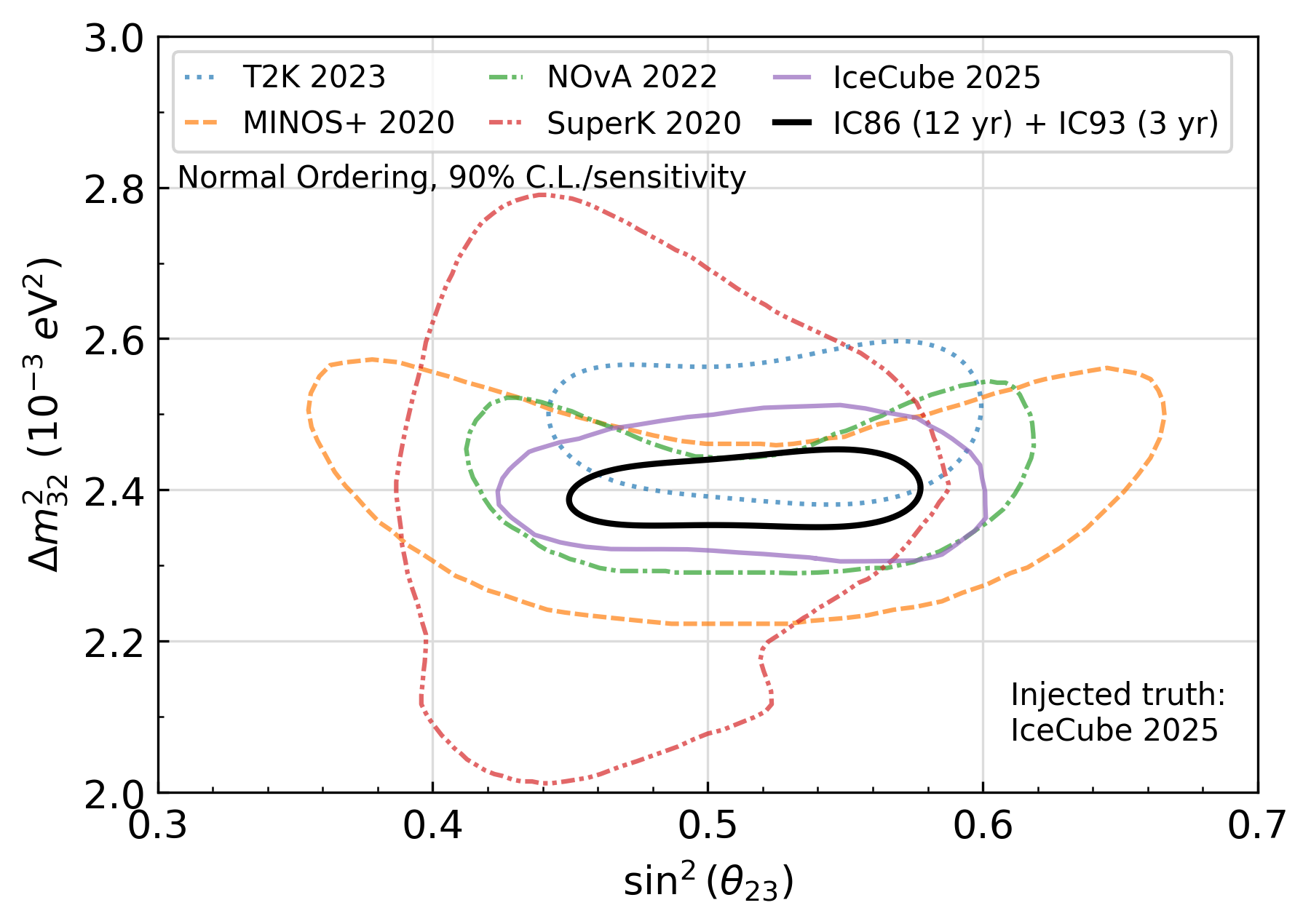}
         }
    \caption{Expected 90\% confidence level contours for the atmospheric oscillation parameters with 3 years of livetime with the new strings, where the assumed truth is given by \cite{IceCube:2024oscnext}.} 
    \label{fig:numu_ic93_vs_other_experiments_FLERCNN}
\end{figure}

In addition to computing the sensitivity for scenarios provided by global fits (Figs. \ref{fig:numu_ic86_vs_ic93} and \ref{fig:numu_ic93_vs_other_experiments}), Figure \ref{fig:numu_ic93_vs_other_experiments_FLERCNN} shows the expected sensitivity for a scenario where the true value realized in nature is the best-fit point as measured by the most recent IceCube DeepCore measurement \cite{IceCube:2024oscnext}.

\section{Reconstruction Comparisons}
\label{sec:appendix_reco}
The analysis sample presented in Section~\ref{sec:data_sample} includes low energy events that would not have been detectable by the existing IC86 detector configuration. Therefore comparing reconstruction resolutions for existing IC86 samples with the new simulated IC93 sample does not offer a direct comparison since the underlying distribution of event types has changed. A more direct comparison of the expected improvement in reconstruction resolutions due to the new strings requires a subset of events from the new IC93 simulated sample that would also have been detected by the IC86 detector configuration.

\begin{figure*}[ht]
     \centering
     \vspace{-1em}
     \subfloat{
         \includegraphics[width=0.6\textwidth]{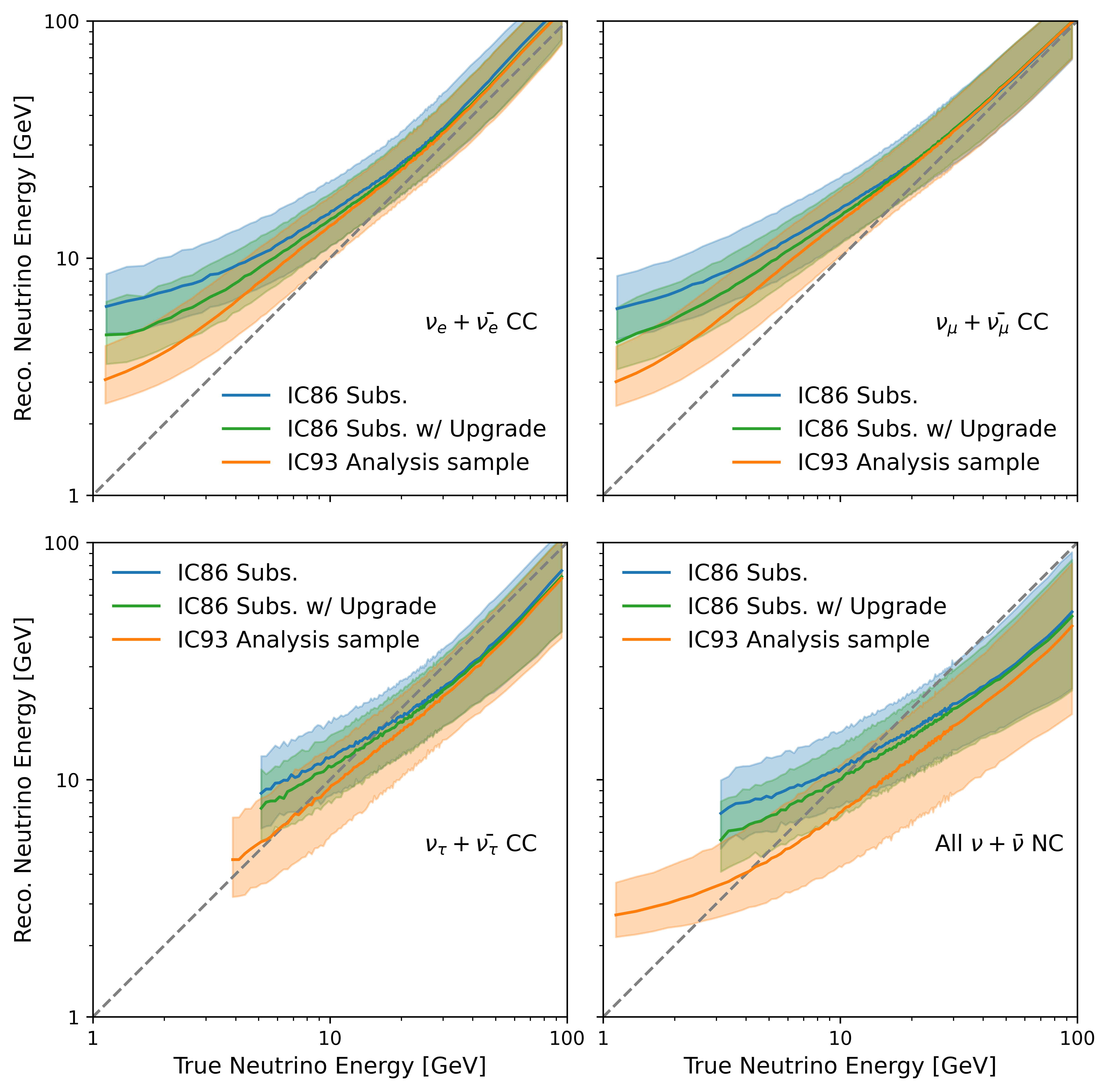}
         }
    \caption{Energy resolution per flavor type for three scenarios: the full IceCube Upgrade analysis sample (orange) and the IC86 subsample with and without the information from the IceCube Upgrade strings (green and blue respectively).} 
    \label{fig:energy_resolution_by_flavor}
\end{figure*}

\begin{figure*}[ht]
     \centering
     \vspace{-1em}
     \subfloat{
         \includegraphics[width=0.6\textwidth]{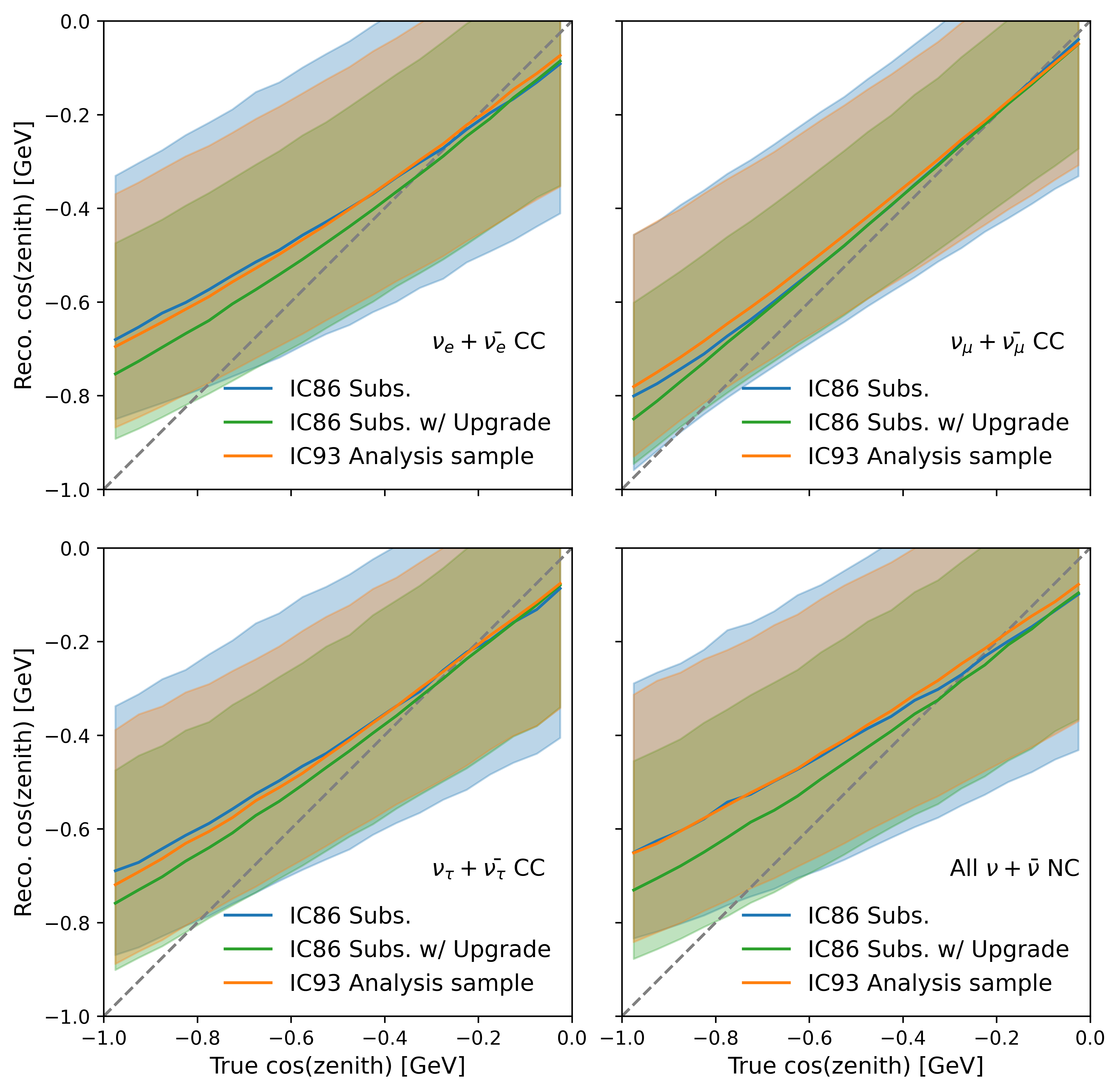}
         }
    \caption{Zenith resolution per flavor type for three scenarios: the full IceCube Upgrade analysis sample (orange) and the IC86 subsample with and without the information from the IceCube Upgrade strings (green and blue respectively).} 
    \label{fig:zenith_resolution_by_flavor}
\end{figure*}

The full sample described in Section~\ref{sec:data_sample} is denoted here as ``IC93 Analysis Sample''. We then identify the subsample of events that are detectable by IC86 and could be improved with the inclusion of the new strings, defined as events with at least 8 pulses in the IC86 array and at least 1 pulse in the IceCube Upgrade extension. This subsample is denoted ``IC86 Subsample''. These events are reconstructed with and without the pulses on the new IceCube Upgrade strings. The reconstruction methods are re-trained for each separate instance. By comparing the performance of the reconstruction methods on the exact same events, with and without the information from the Upgrade strings, we can directly compare the effect of the new hardware on reconstruction performance. Figures \ref{fig:energy_resolution_by_flavor} and \ref{fig:zenith_resolution_by_flavor} quantify the difference in reconstruction for the three scenarios: the full IceCube Upgrade analysis sample (``IC93 Analysis Sample'') and the IC86 subsample with and without the information from the Upgrade strings (``IC86 Subs. w/ Upgrade'' and ``IC86 Subs.'' respectively).

Compared to Fig.~\ref{fig:reco}, these figures are further subdivided by neutrino flavor and interaction channel. In the top panels of  Fig.~\ref{fig:energy_resolution_by_flavor}, it can be observed that $\nu_e + \bar{\nu}_e$ CC and $\nu_\mu + \bar{\nu}_\mu$ CC have both improved variance and bias compared to NC interactions (bottom right of Fig.~\ref{fig:energy_resolution_by_flavor}). This difference is expected as the outgoing neutrino in NC interaction escapes the detector with parts of the energy of the incident neutrino, leading to a systematic underestimation of the energy of the incident neutrino. Similarly, Fig.~\ref{fig:zenith_resolution_by_flavor} contains the angular resolutions shown in Fig.~\ref{fig:reco} subdivided into neutrino flavor and interaction channels. In Fig.~\ref{fig:zenith_resolution_by_flavor}, it can be observed that events in the IC86 subsample have significantly improved variance in zenith reconstructions and improved bias for up-going events for the scenario when information from the IceCube Upgrade strings is included in the reconstruction.

\end{document}